\documentclass[11pt]{article}
\usepackage{graphicx}
\usepackage{amsmath,amssymb,amsthm,amsxtra,overpic,bbm,bm,epsfig,ulem}
\usepackage{color}
\usepackage{multirow}
\usepackage{rotating}
\usepackage{slashbox}
\usepackage{subfigure}

\usepackage{array}
\newcommand{\PreserveBackslash}[1]{\let\temp=\\#1\let\\=\temp}
\newcolumntype{C}[1]{>{\PreserveBackslash\centering}p{#1}}
\newcolumntype{R}[1]{>{\PreserveBackslash\raggedleft}p{#1}}
\newcolumntype{L}[1]{>{\PreserveBackslash\raggedright}p{#1}}

\makeatletter

\newcommand{\Rmnum}[1]{\expandafter\@slowromancap\romannumeral #1@}
\makeatother

\begin{document}

\begin{center}
{\Large \bf On the textures of neutrino mass matrix for
 \\ maximal atmospheric mixing angle and Dirac CP phase}
\end{center}

\vspace{0.05cm}

\begin{center}
{\bf Zhi-Cheng Liu, Chong-Xing Yue and Zhen-hua Zhao \footnote{E-mail: zhzhao@itp.ac.cn}  } \\
{Department of Physics, Liaoning Normal University, Dalian 116029, China}
\end{center}

\vspace{0.2cm}

\begin{abstract}
In this paper, we derive in a novel approach the possible textures of neutrino mass matrix that can lead us to maximal atmospheric mixing angle and Dirac CP phase which are consistent with the current neutrino oscillation data. A total of eleven textures are thus found. Interestingly, the specific texture given by the $\mu$-$\tau$ reflection symmetry can be reproduced from one of the obtained textures. For these textures, some neutrino mass sum rules which relate the neutrino masses and Majorana CP phases will emerge.
\end{abstract}

\newpage

\section{Introduction}

Thanks to the enormous neutrino oscillation data, a framework of three-flavor neutrino mixing has
been established \cite{pdg}. In the basis of charged lepton mass matrix $M^{}_l$ being diagonal, the neutrino mixing matrix $U$ \cite{pmns} originates from diagonalization of the neutrino mass matrix $M^{}_\nu$ in a manner as
\begin{eqnarray}
U^\dagger M^{}_\nu U^* = {\rm Diag}\left( m^{}_1, m^{}_2, m^{}_3 \right) \;,
\label{1}
\end{eqnarray}
with $m^{}_{i}$ (for $i=1, 2, 3$) being the neutrino masses.
In the standard parametrization, $U$ reads
\begin{eqnarray}
U = P^{}_l O^{}_{23} U^{}_{13} O^{}_{12} P^{}_\nu \;,
\label{2}
\end{eqnarray}
where $P^{}_l = {\rm Diag} \left(e^{{\rm i} \phi_e},  e^{{\rm i} \phi_\mu},  e^{{\rm i} \phi_\tau} \right)$ and $P^{}_\nu = {\rm Diag} \left(e^{{\rm i} \rho},  e^{{\rm i} \sigma}, 1 \right)$ are two diagonal phase matrices, and
\begin{eqnarray}
O^{}_{23} = \begin{pmatrix}
1 & 0 & 0  \\ 0 & c^{}_{23} & s^{}_{23} \\ 0 & - s^{}_{23} & c^{}_{23}
\end{pmatrix} \;, \hspace{1cm}
U^{}_{13} = \begin{pmatrix}
c^{}_{13} & 0 & s^{}_{13} e^{-{\rm i}\delta} \\ 0 & 1 & 0 \\ - s^{}_{13} e^{{\rm i}\delta} & 0 & c^{}_{13}
\end{pmatrix} \;, \hspace{1cm}
O^{}_{12} = \begin{pmatrix}
c^{}_{12} & s^{}_{12} & 0 \\ - s^{}_{12} & c^{}_{12} & 0 \\ 0 & 0 & 1
\end{pmatrix} \;,
\label{3}
\end{eqnarray}
with $c^{}_{ij} = \cos{\theta^{}_{ij}}$ and $s^{}_{ij} = \sin{\theta^{}_{ij}}$ for the mixing angles $\theta^{}_{ij}$ (for $ij = 12, 13, 23$). As for the phases,
$\delta$ is known as the Dirac CP phase and responsible for the CP violation effects in neutrino oscillations, while $\rho$ and $\sigma$ are known as the Majorana CP phases and control the rates of  neutrinoless double beta decays that can be used to testify the Majorana nature of neutrinos. And $\phi^{}_{e, \mu, \tau}$ are called unphysical phases since they can be removed by the redefinitions of charged lepton fields. Furthermore, neutrino oscillations are also dependent on the neutrino mass squared differences $\Delta m^{2}_{ij} =  m^2_i - m^2_j$ (for $ij = 21, 31$).

The neutrino oscillation experiments by now give the following results for the neutrino mass squared differences \cite{global}
\begin{eqnarray}
\Delta m^2_{21} = \left(7.50^{+0.19}_{-0.17}\right)\times 10^{-5} \ {\rm eV^2}  \;, \hspace{1cm}
|\Delta m^2_{31}| = (2.524^{+0.039}_{-0.040})\times 10^{-3} \ {\rm eV^2} \;.
\label{4}
\end{eqnarray}
Note that the sign of $\Delta m^2_{31}$ has not yet been determined, thereby allowing for two possible neutrino mass orderings: $m^{}_1 < m^{}_2 < m^{}_3$ (the normal hierarchy and NH for short) and $m^{}_3 < m^{}_1 < m^{}_2$ (the inverted hierarchy and IH for short). Besides, the absolute neutrino mass scale or equivalently the lightest neutrino mass ($m^{}_1$ in the NH case or $m^{}_3$ in the IH case) also remains unknown. On the other hand, the mixing parameters $\theta^{}_{13}$, $\theta^{}_{23}$ and $\delta$ take the values
\begin{eqnarray}
\sin^2{\theta^{}_{13}} = 0.02166 \pm 0.00075  \;, \hspace{1cm}
\sin^2{\theta^{}_{23}} = 0.441 \pm 0.024 \;, \hspace{1cm}
\delta = 261^\circ \pm 55^\circ \;,
\label{5}
\end{eqnarray}
in the NH case, or
\begin{eqnarray}
\sin^2{\theta^{}_{13}} = 0.02179 \pm 0.00076  \;, \hspace{1cm}
\sin^2{\theta^{}_{23}} = 0.587 \pm 0.022 \;, \hspace{1cm}
\delta = 277^\circ \pm 43^\circ \;,
\label{6}
\end{eqnarray}
in the IH case, while $\theta^{}_{12}$ takes the value $\sin^2{\theta^{}_{12}} = 0.306 \pm 0.012$
in either case \cite{global}. But information about $\rho$ and $\sigma$ is still lacking.

It is interesting to note that the current neutrino oscillation data is consistent with maximal atmospheric mixing angle ($\theta^{}_{23} = \pi/4$) and Dirac CP phase ($\delta = - \pi/2$). These remarkable results may point towards some special texture of $M^{}_\nu$. In this regard, the specific texture given by the $\mu$-$\tau$ reflection symmetry \cite{MTR,MTRs} serves as a unique example. This symmetry is defined as follows: In the basis of $M^{}_l$ being diagonal, $M^{}_\nu$ should keep invariant under a combination of the $\mu$-$\tau$ interchange and CP conjugate operations
\begin{eqnarray}
\nu^{}_e \leftrightarrow \nu^c_e \;, \hspace{1cm} \nu^{}_\mu \leftrightarrow \nu^c_\tau \;,
\hspace{1cm} \nu^{}_\tau \leftrightarrow \nu^c_\mu \;,
\label{7}
\end{eqnarray}
and is characterized by
\begin{eqnarray}
M^{}_{e\mu} = M^*_{e\tau} \;, \hspace{1cm} M^{}_{\mu\mu} = M^*_{\tau\tau}  \;, \hspace{1cm}
M^{}_{ee} \ {\rm and} \ M^{}_{\mu\tau} \ {\rm being \ real}  \;,
\label{8}
\end{eqnarray}
where $M^{}_{\alpha \beta}$ denotes the $\alpha \beta$ element of $M^{}_\nu$ (for $\alpha, \beta = e, \mu, \tau$).
Such a texture leads to the following predictions for the neutrino mixing parameters \cite{GL}
\begin{eqnarray}
\phi^{}_e = \frac{\pi}{2} \;, \hspace{1cm} \phi^{}_\mu = - \phi^{}_\tau \;, \hspace{1cm} \theta^{}_{23} = \frac{\pi}{4} \;, \hspace{1cm} \delta = \pm \frac{\pi}{2} \;, \hspace{1cm} \rho, \sigma = 0 \ {\rm or} \ \frac{\pi}{2} \;.
\label{8.1}
\end{eqnarray}
The purpose of this paper is to derive in a novel approach the possible textures of neutrino mass matrix that can give $\theta^{}_{23} = \pi/4$ and $\delta = - \pi/2$ \cite{related}. Such a study may help us reveal the underlying flavor symmetries in the lepton sector. A total of eleven textures are thus found. Interestingly, one of the obtained textures can reproduce the specific texture given by the $\mu$-$\tau$ reflection symmetry. For these textures, some neutrino mass sum rules \cite{sumrule} which relate the neutrino masses and Majorana CP phases will emerge. The rest part of this paper is organized as follows. In section 2, we formulate our approach of deriving the desired textures. In section 3, the derived textures and the resulting neutrino mass sum rules are discussed one by one in some detail. Finally, a summary of our main results is given in section 4.

\section{The approach}

A $3 \times 3$ complex symmetric neutrino mass matrix generally contains twelve degrees of freedom (dfs). After the diagonalization process, three dfs will emerge as the unphysical phases $\phi^{}_{e, \mu, \tau}$ while nine dfs as the physical parameters $\theta^{}_{ij}$, $\delta$, $\rho$, $\sigma$ and $m^{}_{i}$. Therefore, one would suffer some uncertainties due to the unphysical phases when retrodicting the textures of $M^{}_\nu$ based on the characteristics of physical parameters. In comparison, the effective neutrino mass matrix $\bar M^{}_\nu = P^\dagger_l M^{}_\nu P^*_l$ where the unphysical dfs cancel out only consists of nine physical dfs. For this reason, we choose to work on $\bar M^{}_\nu$ instead of $M^{}_\nu$ itself so that the uncertainties due to the unphysical phases can be evaded. Two immediate comments are given as follows: (1) One can recover the results for $M^{}_\nu$ from those for $\bar M^{}_\nu$ by simply making the replacements $\bar M^{}_{\alpha \beta} = M^{}_{\alpha \beta} e^{-{\rm i}(\phi^{}_\alpha+ \phi^{}_\beta)}$ with $\bar M^{}_{\alpha \beta}$ being the $\alpha \beta$ element of $\bar M^{}_\nu$. (2) Since $\bar M^{}_\nu$ only has nine dfs, its twelve components $\bar R^{}_{\alpha \beta} = {\rm Re}(\bar M^{}_{\alpha \beta})$ and
$\bar I^{}_{\alpha \beta} = {\rm Im}(\bar M^{}_{\alpha \beta})$ are not all independent but subject to three constraint equations.

To proceed, we diagonalize $\bar M^{}_\nu$ to give the expressions for the physical parameters in terms of $\bar R^{}_{\alpha \beta} = {\rm Re}(\bar M^{}_{\alpha \beta})$ and
$\bar I^{}_{\alpha \beta} = {\rm Im}(\bar M^{}_{\alpha \beta})$. From Eqs. (\ref{1}-\ref{2}), one gets
\begin{eqnarray}
O^{T}_{12} U^\dagger_{13} O^{T}_{23} \bar M^{}_\nu O^{}_{23} U^*_{13} O^{}_{12} = {\rm Diag}\left( m^{}_1 e^{2{\rm i}\rho}, m^{}_2e^{2{\rm i}\sigma}, m^{}_3 \right) \;.
\label{9}
\end{eqnarray}
In order to simplify the expressions, we define the following three matrices after the rotations $O^{}_{23}$, $U^{}_{13}$ and $O^{}_{12}$ are implemented in succession
\begin{eqnarray}
M^{}_1  = O^{T}_{23} \bar M^{}_\nu O^{}_{23} \;, \hspace{1cm}
M^{}_2 =  U^{\dagger}_{13} M^{}_1 U^{*}_{13} \;, \hspace{1cm}
M^{}_3  =  O^{T}_{12} M^{}_2 O^{}_{12} \;.
\label{10}
\end{eqnarray}
After taking $\theta^{}_{23} = \pi/4$ and $\delta = - \pi/2$, the elements of these three matrices appear as
\begin{eqnarray}
&& M^{11}_1 = \bar M^{}_{ee} \;, \hspace{0.5cm} M^{12}_1 = \displaystyle \frac{\bar M^{}_{e \mu} - \bar M^{}_{e \tau}}{\sqrt 2} \;, \hspace{0.5cm}  M^{13}_1 = \displaystyle \frac{\bar M^{}_{e \mu} + \bar M^{}_{e \tau}}{\sqrt 2} \;,  \nonumber \\
&& M^{22}_1 = \displaystyle \frac{\bar M^{}_{\mu \mu} + \bar M^{}_{\tau \tau}}{2} - \bar M^{}_{\mu \tau} \;, \hspace{0.5cm} M^{23}_1 = \displaystyle \frac{\bar M^{}_{\mu \mu} - \bar M^{}_{\tau \tau}}{2} \;, \hspace{0.5cm}
M^{33}_1 = \displaystyle \frac{\bar M^{}_{\mu \mu} + \bar M^{}_{\tau \tau}}{2} + \bar M^{}_{\mu \tau} \;,
\label{11}
\end{eqnarray}
\begin{eqnarray}
&& M^{11}_2 = c^2_{13} M^{11}_1 - {\rm i} \sin 2 \theta^{}_{13} M^{13}_1 - s^2_{13} M^{33}_1 \;, \hspace{0.5cm}
M^{12}_2 = c^{}_{13} M^{12}_1 - {\rm i} s^{}_{13} M^{23}_1 \;, \nonumber \\
&& M^{13}_2 = \cos 2 \theta^{}_{13} M^{13}_1 - \displaystyle \frac{{\rm i}}{2} \sin 2 \theta^{}_{13} \left( M^{11}_1 + M^{33}_1 \right) \;, \hspace{0.5cm} M^{22}_2 = M^{22}_1 \;, \nonumber \\
&& M^{23}_2 = c^{}_{13} M^{23}_1 - {\rm i} s^{}_{13} M^{12}_1 \;, \hspace{0.5cm}
M^{33}_2 = c^2_{13} M^{33}_1 - {\rm i} \sin 2 \theta^{}_{13} M^{13}_1 - s^2_{13} M^{11}_1 \;,
\label{12}
\end{eqnarray}
and
\begin{eqnarray}
&& M^{11}_3 = c^2_{12} M^{11}_2 - \sin 2 \theta^{}_{12} M^{12}_2 + s^2_{12} M^{22}_2 \;, \nonumber \\
&& M^{12}_3 = \cos 2 \theta^{}_{12} M^{12}_2 + \displaystyle \frac{1}{2} \sin 2 \theta^{}_{12} \left( M^{11}_2 - M^{22}_{2} \right) \;, \hspace{0.5cm} M^{13}_3 = c^{}_{12} M^{13}_2 - s^{}_{12} M^{23}_2 \;,  \nonumber \\ && M^{22}_3 = s^2_{12} M^{11}_2 + \sin 2 \theta^{}_{12} M^{12}_2 + c^2_{12} M^{22}_2 \;,  \hspace{0.5cm}  M^{23}_3 = s^{}_{12} M^{13}_2 + c^{}_{12} M^{23}_2 \;, \hspace{0.5cm}  M^{33}_3 = M^{33}_2 \;,
\label{13}
\end{eqnarray}
where $M^{ij}_{k}$ stands for the $ij$ element of $M^{}_k$ (for $i,j,k=1,2,3$). In order for the diagonalization process to work, one must have
\begin{eqnarray}
{\rm Re/Im} \left( M^{13}_2 \right) = {\rm Re/Im} \left( M^{23}_2 \right) = {\rm Re/Im} \left( M^{12}_3 \right) = {\rm Im} \left( M^{33}_3 \right) = 0 \;,
\label{14}
\end{eqnarray}
and
\begin{eqnarray}
m^{}_1 e^{2{\rm i}\rho} = M^{11}_{3} \;, \hspace{1cm} m^{}_2 e^{2{\rm i}\sigma} = M^{22}_{3} \;, \hspace{1cm} m^{}_3 = M^{33}_{3} \;.
\label{15}
\end{eqnarray}
The seven conditions in Eq. (\ref{14}) which will be referred to as A-G in order are explicitly written as
\begin{eqnarray}
&& {\rm A:} \quad 2 \cos 2 \theta^{}_{13} R^{13}_1 = - \sin 2 \theta^{}_{13} \left( I^{11}_1 + I^{33}_1 \right) \;, \nonumber \\
&& {\rm B:} \quad 2 \cos 2 \theta^{}_{13} I^{13}_1 =  \sin 2 \theta^{}_{13} \left( R^{11}_1 + R^{33}_1 \right) \;, \nonumber \\
&& {\rm C:} \quad c^{}_{13} R^{23}_1 = - s^{}_{13} I^{12}_1 \;, \nonumber \\
&& {\rm D:} \quad c^{}_{13} I^{23}_1 =   s^{}_{13} R^{12}_1 \;, \nonumber \\
&& {\rm E:} \quad 2 \cos 2 \theta^{}_{12} R^{12}_2 = - \sin 2 \theta^{}_{12} \left( R^{11}_2 - R^{22}_{2} \right) \;, \nonumber \\
&& {\rm F:} \quad 2 \cos 2 \theta^{}_{12} I^{12}_2 = - \sin 2 \theta^{}_{12} \left( I^{11}_2 - I^{22}_{2} \right) \;, \nonumber \\
&& {\rm G:} \quad \sin 2 \theta^{}_{13} R^{13}_1 = c^2_{13} I^{33}_1 - s^2_{13} I^{11}_1 \;,
\label{16}
\end{eqnarray}
with $R^{ij}_{k} = {\rm Re} \left( M^{ij}_{k} \right)$ and $I^{ij}_{k} = {\rm Im} \left( M^{ij}_{k} \right)$.

The expressions for $\theta^{}_{12}$ and $\theta^{}_{13}$ in terms of the components of $\bar M^{}_\nu$ can be directly read from Eq. (\ref{16}). For example, equation E gives $\theta^{}_{12}$ as
\begin{eqnarray}
\tan 2 \theta^{}_{12} = \frac{-2 R^{12}_2}{ R^{11}_2 - R^{22}_{2}} \;,
\label{17}
\end{eqnarray}
while equation F gives $\theta^{}_{12}$ as
\begin{eqnarray}
\tan 2 \theta^{}_{12} = \frac{-2 I^{12}_2}{ I^{11}_2 - I^{22}_{2}} \;.
\label{18}
\end{eqnarray}
By relating these two expressions for $\theta^{}_{12}$, a constraint equation for the components of $\bar M^{}_\nu$ arises as
\begin{eqnarray}
&& {\rm EF}: \quad R^{12}_2 \left( I^{11}_2 - I^{22}_{2} \right) = I^{12}_2 \left( R^{11}_2 - R^{22}_{2} \right) \;,
\label{19}
\end{eqnarray}
where the symbol EF is used to indicate that this constraint equation results from equations E and F.
It can be expressed in terms of $\bar R^{}_{\alpha \beta} $ and $\bar I^{}_{\alpha \beta}$ by taking the expressions
\begin{eqnarray}
R^{12}_2  &=& {\rm sgn} \left(\bar R^{}_{e \mu} - \bar R^{}_{e \tau}\right) \sqrt{\frac{1}{2} \left(\bar R^{}_{e \mu} - \bar R^{}_{e \tau}\right)^2 + \frac{1}{4} \left( \bar I^{}_{\mu \mu} - \bar I^{}_{\tau \tau} \right)^2 } \;, \nonumber  \\
I^{12}_2
&=& {\rm sgn} \left(\bar I^{}_{e \mu} - \bar I^{}_{e \tau}\right) \sqrt{\frac{1}{2} \left(\bar I^{}_{e \mu} - \bar I^{}_{e \tau}\right)^2 + \frac{1}{4} \left( \bar R^{}_{\mu \mu} - \bar R^{}_{\tau \tau} \right)^2 } \;, \nonumber \\
I^{11}_2 - I^{22}_2 & = & \bar I^{}_{ee} - \bar I^{}_{\mu\mu} - \bar I^{}_{\tau\tau} \;, \nonumber \\
R^{11}_2 - R^{22}_2  &=& \frac{\bar R^{}_{ee} + \bar R^{}_{\mu \tau}}{2} - \frac{3}{4} \left(\bar R^{}_{\mu \mu} + \bar R^{}_{\tau \tau}\right) + {\rm sgn} \left(\bar I^{}_{e \mu} + \bar I^{}_{e \tau}\right) \nonumber \\
&& \times \sqrt{\frac{1}{2} \left( \bar I^{}_{e \mu} + \bar I^{}_{e \tau}\right)^2 + \frac{1}{4} \left(\bar R^{}_{ee} + \bar R^{}_{\mu \tau} + \frac{ \bar R^{}_{\mu \mu} + \bar R^{}_{\tau \tau}}{2}\right)^2 } \;.
\label{20}
\end{eqnarray}
In a similar way, one will arrive at the following constraint equations by relating the expressions for $\theta^{}_{13}$ derived from equations A-D
\begin{eqnarray}
&& {\rm AB}: \quad R^{13}_1 \left( R^{11}_1 + R^{33}_1 \right) = - I^{13}_1 \left( I^{11}_1 + I^{33}_1 \right) \;, \nonumber \\
&& {\rm AC}: \quad I^{12}_1 R^{23}_1 \left( I^{11}_1 + I^{33}_1 \right) = R^{13}_1 \left[ \left(I^{12}_1\right)^2 - \left( R^{23}_1\right)^2 \right] \;, \nonumber \\
&& {\rm AD}: \quad I^{23}_1 R^{12}_1 \left( I^{11}_1 + I^{33}_1 \right) = - R^{13}_1 \left[ \left(R^{12}_1\right)^2 - \left( I^{23}_1 \right)^2 \right] \;, \nonumber \\
&& {\rm BC}: \quad - I^{12}_1 R^{23}_1 \left( R^{11}_1 + R^{33}_1 \right) =  I^{13}_1 \left[ \left(I^{12}_1\right)^2 - \left( R^{23}_1\right)^2 \right] \;, \nonumber \\
&& {\rm BD}: \quad I^{23}_1 R^{12}_1 \left( R^{11}_1 + R^{33}_1 \right) = I^{13}_1 \left[ \left(R^{12}_1\right)^2 - \left( I^{23}_1 \right)^2 \right] \;, \nonumber \\
&& {\rm CD}: \quad R^{23}_1 R^{12}_1 = - I^{23}_1 I^{12}_1 \;,
\label{21}
\end{eqnarray}
which in terms of $\bar R^{}_{\alpha \beta} $ and $\bar I^{}_{\alpha \beta}$ are respectively expressed as
\begin{eqnarray}
&& \hspace{-1cm} \left(\bar R_{e \mu} + \bar R_{e \tau}\right) \left(\bar R_{ee} + \bar R_{\mu \tau} + \frac{\bar R_{\mu \mu} + \bar R_{\tau \tau}}{2} \right)
= - \left(\bar I_{e \mu} + \bar I_{e \tau}\right) \left(\bar I_{ee} + \bar I_{\mu \tau} + \frac{\bar I_{\mu \mu} + \bar I_{\tau \tau}}{2} \right) \;, \nonumber \\
&& \hspace{-1cm} \left(\bar I_{e \mu} - \bar I_{e \tau}\right) \left(\bar R_{\mu \mu} - \bar R_{\tau \tau}\right)
\left(\bar I_{ee} + \bar I_{\mu \tau} + \frac{\bar I_{\mu \mu} + \bar I_{\tau \tau}}{2} \right)
= \left(\bar R_{e \mu} + \bar R_{e \tau}\right) \Big[ \left(\bar I_{e \mu} - \bar I_{e \tau}\right)^2  \nonumber \\
&& \hspace{7.8cm} - \frac{1}{2} \left(\bar R_{\mu \mu} - \bar R_{\tau \tau}\right)^2 \Big]
\;, \nonumber \\
&& \hspace{-1cm} \left( \bar I_{\mu \mu} - \bar I_{\tau \tau} \right) \left(\bar R_{e \mu} - \bar R_{e \tau}\right)
\left(\bar I_{ee} + \bar I_{\mu \tau} + \frac{\bar I_{\mu \mu} + \bar I_{\tau \tau}}{2} \right)
= - \left(\bar R_{e \mu} + \bar R_{e \tau}\right) \Big[ \left( \bar R_{e \mu} - \bar R_{e \tau} \right)^2 \nonumber \\
&& \hspace{7.8cm} - \frac{1}{2} \left( \bar I_{\mu \mu} - \bar I_{\tau \tau} \right)^2 \Big] \;, \nonumber \\
&& \hspace{-1cm} \left(\bar I_{e \mu} - \bar I_{e \tau}\right) \left(\bar R_{\mu \mu} - \bar R_{\tau \tau}\right)
\left(\bar R_{ee} + \bar R_{\mu \tau} + \frac{\bar R_{\mu \mu} + \bar R_{\tau \tau}}{2} \right)
= - \left(\bar I_{e \mu} + \bar I_{e \tau}\right) \Big[ \left(\bar I_{e \mu} - \bar I_{e \tau}\right)^2 \nonumber \\
&& \hspace{8.3cm} - \frac{1}{2} \left(\bar R_{\mu \mu} - \bar R_{\tau \tau}\right)^2 \Big]
\;, \nonumber \\
&& \hspace{-1cm} \left( \bar I_{\mu \mu} - \bar I_{\tau \tau} \right) \left(\bar R_{e \mu} - \bar R_{e \tau}\right)
\left(\bar R_{ee} + \bar R_{\mu \tau} + \frac{\bar R_{\mu \mu} + \bar R_{\tau \tau}}{2} \right)
= \left(\bar I_{e \mu} + \bar I_{e \tau}\right) \Big[ \left( \bar R_{e \mu} - \bar R_{e \tau} \right)^2 \nonumber \\
&& \hspace{8.3cm} - \frac{1}{2} \left( \bar I_{\mu \mu} - \bar I_{\tau \tau} \right)^2 \Big]
\;, \nonumber \\
&& \hspace{-1cm} \left(\bar R_{\mu \mu} - \bar R_{\tau \tau}\right) \left(\bar R_{e \mu} - \bar R_{e \tau}\right)
= - \left(\bar I_{\mu \mu} - \bar I_{\tau \tau}\right) \left(\bar I_{e \mu} - \bar I_{e \tau}\right) \;.
\label{22}
\end{eqnarray}
But not all of these six constraint equations are independent. For example, equation BC can be derived from equations AB and AC. In fact, at most three of them can be independent. A set of three independent constraint equations (e.g., AB, AC and AD) can be chosen in such a way that each of equations A-D has been used at least once in deriving them. Finally, we obtain a constraint equation as
\begin{eqnarray}
&& {\rm AG}: \quad - I^{11}_1 + I^{33}_1 = {\rm sgn} \left(R^{13}_1\right) \sqrt{4 \left(R^{13}_1\right)^2 + \left(I^{11}_1 + I^{33}_1\right)^2} \;,
\label{23}
\end{eqnarray}
by relating the expressions for $\theta^{}_{13}$ derived from equations A and G.
Its expression in terms of $\bar R^{}_{\alpha \beta} $ and $\bar I^{}_{\alpha \beta}$ appears as
\begin{eqnarray}
\hspace{-1cm} - \bar I^{}_{ee} + \bar I^{}_{\mu \tau} + \frac{\bar I^{}_{\mu \mu} + \bar I^{}_{\tau \tau}}{2}  ={\rm sgn} \left(\bar R^{}_{e \mu} + \bar R^{}_{e \tau}\right) \sqrt{2 \left(\bar R^{}_{e \mu} + \bar R^{}_{e \tau}\right)^2 +  \left(\bar I^{}_{ee} + \bar I^{}_{\mu \tau} + \frac{\bar I^{}_{\mu \mu} + \bar I^{}_{\tau \tau}}{2}  \right)^2} \;.
\label{24}
\end{eqnarray}
To sum up, a total of five independent constraint equations for the components of $\bar M^{}_\nu$ (i.e., Eqs. (\ref{20}, \ref{24}) and three independent ones from Eq. (\ref{22})) will arise from the eliminations of $\theta^{}_{12}$ and $\theta^{}_{13}$ in Eq. (\ref{16}). This can be understood from the fact that two more conditions (i.e., $\theta^{}_{23} = \pi/4$ and $\delta = -\pi/2$) have been imposed on the basis of three intrinsic constraint equations for the components of $\bar M^{}_\nu$. At last, one can say that an $\bar M^{}_\nu$ with its components satisfying these constraint equations will necessarily produce $\theta^{}_{23} = \pi/4$ and $\delta = -\pi/2$.

By taking the expressions for $\theta^{}_{12}$ and $\theta^{}_{13}$ derived from Eq. (\ref{16}) in Eq. (\ref{15}), the neutrino masses in combination with the Majorana CP phases are expressed as
\begin{eqnarray}
{\rm Re} \left(m^{}_1 e^{2{\rm i}\rho}\right) &=& \frac{R^{11}_2 + R^{22}_2}{2}
- {\rm sgn} \left(R^{12}_2\right) \sqrt{\left(R^{12}_2\right)^2 + \frac{1}{4}\left(R^{11}_2 - R^{22}_2\right)^2 } \;, \nonumber \\
{\rm Im} \left(m^{}_1 e^{2{\rm i}\rho}\right) &=& \frac{I^{11}_2 + I^{22}_2}{2}
- {\rm sgn} \left(I^{12}_2\right) \sqrt{ \left(I^{12}_2\right)^2 + \frac{1}{4} \left(I^{11}_2 - I^{22}_2\right)^2 } \;, \nonumber \\
{\rm Re} \left(m^{}_2 e^{2{\rm i}\sigma}\right) &=& \frac{R^{11}_2 + R^{22}_2}{2}
+ {\rm sgn} \left(R^{12}_2\right) \sqrt{ \left(R^{12}_2\right)^2 + \frac{1}{4} \left(R^{11}_2 - R^{22}_2\right)^2 } \;, \nonumber \\
{\rm Im} \left(m^{}_2 e^{2{\rm i}\sigma}\right) &=& \frac{I^{11}_2 + I^{22}_2}{2}
+ {\rm sgn} \left(I^{12}_2\right) \sqrt{ \left(I^{12}_2\right)^2 + \frac{1}{4} \left(I^{11}_2 - I^{22}_2\right)^2 } \;, \nonumber \\
m^{}_3 &=& - \frac{\bar R^{}_{ee} - \bar R^{}_{\mu \tau}}{2} + \frac{\bar R^{}_{\mu \mu} + \bar R^{}_{\tau \tau}}{4} + {\rm sgn} \left(\bar I^{}_{e \mu} + \bar I^{}_{e \tau}\right) \nonumber \\
&&  \times \sqrt{\frac{1}{2}\left(\bar I^{}_{e \mu} + \bar I^{}_{e \tau}\right)^2 + \frac{1}{4} \left(\bar R^{}_{ee} + \bar R^{}_{\mu \tau} + \frac{\bar R^{}_{\mu \mu} + \bar R^{}_{\tau \tau}}{2} \right)^2}\;,
\label{25}
\end{eqnarray}
where
\begin{eqnarray}
R^{11}_2 + R^{22}_2  &=& \frac{\bar R^{}_{ee} - 3\bar R^{}_{\mu \tau}}{2} + \frac{\bar R^{}_{\mu \mu} + \bar R^{}_{\tau \tau}}{4} + {\rm sgn} \left(\bar I^{}_{e \mu} + \bar I^{}_{e \tau}\right) \nonumber \\
&& \times \sqrt{\frac{1}{2} \left( \bar I^{}_{e \mu} + \bar I^{}_{e \tau}\right)^2 + \frac{1}{4} \left(\bar R^{}_{ee} + \bar R^{}_{\mu \tau} + \frac{\bar R^{}_{\mu \mu} + \bar R^{}_{\tau \tau}}{2} \right)^2 }\;, \nonumber \\
I^{11}_2 + I^{22}_2
&=&  \bar I^{}_{ee} - 2 \bar I^{}_{\mu\tau} \;,
\label{26}
\end{eqnarray}
while the expressions for $R^{12}_{2}, I^{12}_{2}, R^{11}_2-  R^{22}_2$ and $R^{11}_2-  R^{22}_2$ have been given in Eq. (\ref{20}).

We point out that the above results are derived in the general case where none of equations A-G has its two sides vanish. However, there is the possibility that $\bar M^{}_\nu$ has such a special texture that one (or more) of equations A-G has its two sides vanish and thus always holds irrespective of the values of $\theta^{}_{12}$ and $\theta^{}_{13}$. Since the purpose of this paper is to derive the possible textures of neutrino mass matrix that can lead to $\theta^{}_{23} =\pi/4$ and $\delta = -\pi/2$, such a possibility should of course be taken into account. When an equation has its two sides vanish, it fails to give an expression for $\theta^{}_{12}$ or $\theta^{}_{13}$ and thus the constraint equation(s) resulting from it will become ineffective. But the fact that its two sides vanish itself will bring about two new constraint equations. It turns out that in such kind of case the number of independent constraint equations will get increased compared to in the general case. (For example, in the case of two sides of equation E being vanishing, the expression for $\theta^{}_{12}$ in Eq. (\ref{17}) and thus equation EF become ineffective. But there are two new constraint equations $R^{12}_2 = R^{11}_2 - R^{22}_{2} =0$. So, in effect, the number of independent constraint equations in this case gets increased by one compared to in the general case.)
When this number gets increased by one (and so on), there will correspondingly be one (and so on) neutrino mass sum rule as we will see. In the next section, all the possible cases where one or more of equations A-G have their two sides vanish will be examined. Before doing that, we make a few observations: (1) In the case of two sides of equation A being vanishing (i.e., $R^{13}_{1} = I^{11}_{1} + I^{33}_{1} = 0$), two sides of equation G are necessarily also vanishing (i.e., $R^{13}_{1} = c^2_{13} I^{33}_1 - s^2_{13} I^{11}_1= 0$). The reverse is also true. It is easy to see that a combination of $c^2_{13} I^{33}_1 - s^2_{13} I^{11}_1 = 0$ and $I^{11}_{1} + I^{33}_{1} = 0$ leads us to $I^{11}_{1} = I^{33}_{1} = 0$. So equations A and G always have their two sides vanish simultaneously. And in such a case one has
\begin{eqnarray}
R^{13}_1 = I^{11}_1 = I^{33}_1 = 0 \;.
\label{27}
\end{eqnarray}
(2) In the case of two sides of equation C being vanishing (i.e., $R^{23}_1 = I^{12}_1 = 0$), as a result of the relation $I^{12}_2 = c^{}_{13} I^{12}_1 - s^{}_{13} R^{23}_1$, two sides of equation F are also vanishing (i.e., $I^{12}_2 = I^{11}_2 - I^{22}_{2} =0$). The reverse is also true. So equations C and F always have their two sides vanish simultaneously. And in such a case one has
\begin{eqnarray}
I^{11}_2 - I^{22}_{2} = R^{23}_1 = I^{12}_1 =  0 \;.
\label{28}
\end{eqnarray}
(3) In the case of two sides of equation D being vanishing (i.e., $I^{23}_1 = R^{12}_1 = 0$), as a result of the relation $R^{12}_2 = c^{}_{13} R^{12}_1 + s^{}_{13} I^{23}_1$, two sides of equation E are also vanishing (i.e., $R^{12}_2 = R^{11}_2 - R^{22}_{2} =0$). The reverse is also true. So equations D and E always have their two sides vanish simultaneously. And in such a case one has
\begin{eqnarray}
R^{11}_2 - R^{22}_{2}= I^{23}_1 = R^{12}_1 =  0 \;.
\label{29}
\end{eqnarray}
(4) Equations E and F (or A, B, C, D and G) are not allowed to have their two sides vanish simultaneously. Otherwise, $\theta^{}_{12}$ (or $\theta^{}_{13}$) would be free of any constraint and have no reason to take the measured value. For these observations, we just need to consider the cases where
equations A\&G, B, C\&F, D\&E, A\&B\&G, A\&C\&F\&G, A\&D\&E\&G, B\&C\&F, B\&D\&E, A\&B\&C\&F\&G or A\&B\&D\&E\&G have their two sides vanish.

\section{Various textures}

For later use, we give the expressions for the elements of $\bar M^{}_\nu$ in terms of the physical parameters	
\begin{eqnarray}
\bar M^{}_{ee} & = &  m^{}_1 e^{2 {\rm i} \rho } c^2_{12} c^2_{13} + m^{}_2 e^{2 {\rm i} \sigma } s^2_{12} c^2_{13} - m^{}_3 s^2_{13} \;, \nonumber \\
\bar M^{}_{e\mu} & = & \frac{1}{\sqrt 2} \left[ m^{}_1 e^{2 {\rm i} \rho } c^{}_{12} \left(- s^{}_{12} + {\rm i} c^{}_{12} s^{}_{13}\right)
+ m^{}_2 e^{2 {\rm i} \sigma } s^{}_{12} \left(c^{}_{12}+ {\rm i} s^{}_{12} s^{}_{13}\right) + {\rm i} m^{}_3 s^{}_{13} \right] c^{}_{13} \;, \nonumber \\
\bar M^{}_{e\tau} & = & \frac{1}{\sqrt 2} \left[ m^{}_1 e^{2 {\rm i} \rho } c^{}_{12} \left(s^{}_{12}+ {\rm i} c^{}_{12} s^{}_{13}\right)
+ m^{}_2 e^{2 {\rm i} \sigma } s^{}_{12} \left( - c^{}_{12} + {\rm i} s^{}_{12} s^{}_{13}\right) + {\rm i} m^{}_3 s^{}_{13} \right] c^{}_{13} \;, \nonumber \\
\bar M^{}_{\mu\mu} & = & \frac{1}{2} \left[ m^{}_1 e^{2 {\rm i} \rho } \left(s^{}_{12}-{\rm i} c^{}_{12} s^{}_{13}\right)^2
+ m^{}_2 e^{2 {\rm i} \sigma } \left(c^{}_{12}+ {\rm i} s^{}_{12} s^{}_{13}\right)^2 + m^{}_3 c^2_{13}  \right] \;, \nonumber \\
\bar M^{}_{\mu\tau} & = & \frac{1}{2} \left[ - m^{}_1 e^{2 {\rm i} \rho } \left(s^2_{12} + c^2_{12} s^2_{13} \right) - m^{}_2 e^{2 {\rm i} \sigma } \left(c^2_{12}+s^2_{12} s^2_{13}\right) + m^{}_3 c^2_{13}  \right] \;, \nonumber \\
\bar M^{}_{\tau\tau} & = & \frac{1}{2} \left[ m^{}_1 e^{2 {\rm i} \rho } \left(s^{}_{12}+{\rm i} c^{}_{12} s^{}_{13}\right)^2
+ m^{}_2 e^{2 {\rm i} \sigma } \left(c^{}_{12}-{\rm i} s^{}_{12} s^{}_{13}\right)^2 + m^{}_3 c^2_{13} \right] \;,
\label{29.1}
\end{eqnarray}
which are obtained in the way as
\begin{eqnarray}
&& \bar M^{}_\nu = O^{}_{23}  U^{}_{13} O^{}_{12}  {\rm Diag}\left( m^{}_1 e^{2{\rm i}\rho}, m^{}_2e^{2{\rm i}\sigma}, m^{}_3 \right) O^{T}_{12}  U^{T}_{13} O^{T}_{23} \;.
\label{29.2}
\end{eqnarray}
In the calculations, $\theta^{}_{23} = \pi/4$ and $\delta  = - \pi/2$ have been input.

\subsection{A\&G}

In the case of two sides of equations A and G being vanishing, equations AB, AC, AD and AG become ineffective. We are left with three independent constraint equations (i.e., equation EF and two independent ones of equations BC, BD and CD). But, as discussed at the end of section 2, there are three new constraint equations which are given by Eq. (\ref{27}) and lead to the following relations for $\bar R^{}_{\alpha \beta}$ and $\bar I^{}_{\alpha \beta}$
\begin{eqnarray}
\bar R^{}_{e \mu} = -\bar R^{}_{e \tau} \;, \hspace{1cm} \bar I^{}_{ee} = 0 \;, \hspace{1cm} -2 \bar I^{}_{\mu \tau} = \bar I^{}_{\mu \mu} + \bar I^{}_{\tau \tau} \;.
\label{30}
\end{eqnarray}
By taking these relations, the expressions for the surviving constraint equations  can be simplified to some extent. In total, the number of independent constraint equations  gets increased by one compared to in the general case. So one neutrino mass sum rule will arise.

With the help of Eq. (\ref{29.1}), one can easily get the desired neutrino mass sum rule as
\begin{eqnarray}
m^{}_1 c^2_{12} \sin 2 \rho + m^{}_2 s^2_{12} \sin 2 \sigma = 0 \;,
\label{32}
\end{eqnarray}
from any one of the relations in Eq. (\ref{30}). This can be verified by taking the relations in Eq. (\ref{30}) in the expressions for the neutrino masses in combination with the Majorana CP phases in Eq. (\ref{25}).
We discuss the implications of this sum rule in three cases:
(1) In the case of $\sin 2\rho = \sin 2\sigma = 0$, the neutrino masses become independent of the Majorana CP phases. As a result of ${\rm Im}(m^{}_1 e^{2 {\rm i} \rho}) = {\rm Im}(m^{}_2 e^{2 {\rm i} \sigma}) = 0$, an additional constraint equation
\begin{eqnarray}
\bar I^{}_{\mu \tau} = 0 \;,
\label{33}
\end{eqnarray}
arises from Eq. (\ref{25}) under the condition of Eq. (\ref{30}). It is found that Eq. (\ref{33}) together with Eq. (\ref{30}) will lead us to a situation where two sides of equations C and F are also vanishing. This happens to be the case of two sides of equations A, C, F and G being vanishing which will be discussed in subsection 3.6.
(2) In the case of $m^{}_1 = 0$, one immediately obtains $\sigma = 0$ or $\pi/2$.
As a result of ${\rm Re}(m^{}_1 e^{2 {\rm i} \rho}) = {\rm Im}(m^{}_1 e^{2 {\rm i} \rho}) = {\rm Im}(m^{}_2 e^{2 {\rm i} \sigma}) = 0$, two additional constraint equations
arise from Eq. (\ref{25}) under the condition of Eq. (\ref{30}). In this case, the effective neutrino mass
\begin{eqnarray}
\left| \langle m \rangle^{}_{ee} \right| = \left| m^{}_1 e^{2{\rm i}\rho} c^2_{12} c^2_{13} + m^{}_2 e^{2{\rm i}\sigma} s^2_{12} c^2_{13} + m^{}_3 s^2_{13} e^{-2{\rm i}\delta} \right| \;,
\label{40}
\end{eqnarray}
which controls the rates of neutrinoless double beta decays takes a value of $m^{}_2 s^2_{12} c^2_{13}- m^{}_3 s^2_{13} \simeq 0.002$ eV for $\sigma = 0$ or $m^{}_2 s^2_{12} c^2_{13} + m^{}_3 s^2_{13} \simeq 0.004$ eV for $\sigma = \pi/2$.
(3) In the case of $\sin 2\rho$ and $\sin 2\sigma \neq 0$, we present $\sin2\sigma/\sin2\rho$ as a function of the lightest neutrino mass in the NH and IH cases in Fig. 1. In the IH case, it takes a value close to $-c^2_{12}/s^2_{12} \simeq -2.27$ in the whole mass range as a result of $m^{}_1 \simeq m^{}_2$. In the NH case, the value is very small for vanishingly small $m^{}_1$ but approaches $-2.27$ for $m^{}_1 \simeq m^{}_2 \simeq 0.1$ eV.

\begin{figure}
\centering
\includegraphics[width=0.45\textwidth]{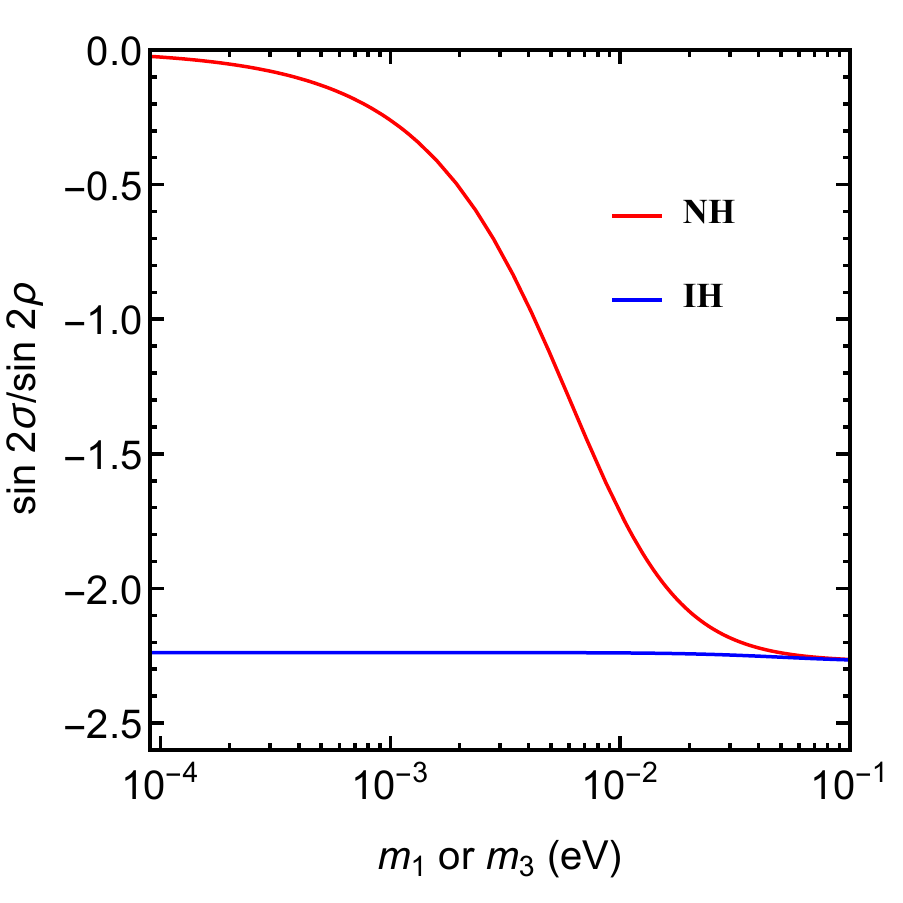}
\caption{$\sin2\sigma/\sin2\rho$ as a function of the lightest neutrino mass ($m^{}_1$ in the NH case or $m^{}_3$ in the IH case) in the case of two sides of equations A and G being vanishing.}
\end{figure}

\subsection{B}

In the case of two sides of equation B being vanishing, equations AB, BC and BD become ineffective. We are left with four independent constraint equations (i.e., equations AG, EF and two independent ones of equations AC, AD and CD). But there are two new constraint equations
\begin{eqnarray}
I^{13}_1 = R^{11}_1 + R^{33}_1 = 0 \;,
\label{34}
\end{eqnarray}
which lead to the following relations for $\bar R^{}_{\alpha \beta} $ and $\bar I^{}_{\alpha \beta}$
\begin{eqnarray}
\bar I^{}_{e \mu} = - \bar I^{}_{e \tau} \;, \hspace{1cm} -2 \left( \bar R^{}_{ee} + \bar R^{}_{\mu \tau} \right) =  \bar R^{}_{\mu \mu} + \bar R^{}_{\tau \tau} \;.
\label{35}
\end{eqnarray}
By taking these relations, the expressions for the surviving constraint equations  can be simplified to some extent. In total, the number of independent constraint equations  gets increased by one compared to in the general case. So one neutrino mass sum rule will arise.

With the help of Eq. (\ref{29.1}), one can easily get the desired neutrino mass sum rule as
\begin{eqnarray}
m^{}_1 c^2_{12} \cos 2 \rho + m^{}_2 s^2_{12} \cos 2 \sigma + m^{}_3 = 0 \;,
\label{37}
\end{eqnarray}
from any one of the relations in Eq. (\ref{35}). This can be verified by taking the relations in Eq. (\ref{35}) in the expressions for the neutrino masses in combination with the Majorana CP phases in Eq. (\ref{25}).
With the help of the inequality
\begin{eqnarray}
m^{}_1 c^2_{12} \cos 2 \rho + m^{}_2 s^2_{12} \cos 2 \sigma
\ge - m^{}_1 c^2_{12}  - m^{}_2 s^2_{12} > - m^{}_2 \;,
\label{38}
\end{eqnarray}
one can see that this sum rule can never be fulfilled in the NH case.
In the case of $\cos 2 \rho =0$ (or $\cos 2 \sigma =0$) where an additional constraint equation as
\begin{eqnarray}
- \bar R_{\mu\tau} \ ( {\rm or} \  \bar R_{\mu\tau} ) = {\rm sgn} \left(\bar R_{e \mu} - \bar R_{e \tau}\right)
\sqrt{\frac{1}{2} \left(\bar R_{e \mu} - \bar R_{e \tau}\right)^2 + \frac{1}{4} \left( \bar I_{\mu \mu} - \bar I_{\tau \tau} \right)^2 + \left(\bar R_{ee} + \bar R_{\mu\tau}\right)^2 } \;,
\label{41}
\end{eqnarray}
arises from Eq. (\ref{25}) under the condition of Eq. (\ref{35}),
$\sigma$ (or $\rho$) and $\left| \langle m \rangle^{}_{ee} \right|$ are presented as functions of $m^{}_3$ in Fig. 2. For vanishingly small $m^{}_3$, $\sigma$ (or $\rho$) takes a value close to $\pi/4$ or $3\pi/4$ while $\left| \langle m \rangle^{}_{ee} \right|$ takes a value close to $m^{}_1 c^2_{12} c^2_{13} + m^{}_2 s^2_{12} c^2_{13} \simeq 0.049$ eV for $\rho \simeq \sigma$ or $m^{}_1 c^2_{12} c^2_{13} - m^{}_2 s^2_{12} c^2_{13} \simeq 0.019$ eV for $\rho \simeq \sigma + \pi/2$. When $m^{}_3$ takes the upper value $m^{}_2 s^2_{12} \simeq 0.016$ eV (or $m^{}_1 c^2_{12} \simeq 0.048$ eV), $\sigma$ (or $\rho$) becomes $\pi/2$ while $\left| \langle m \rangle^{}_{ee} \right|$ becomes $\sqrt{\left(m^{}_1 c^2_{12} c^2_{13}\right)^2 + \left(m^{}_2 s^2_{12} c^2_{13} + m^{}_3 s^2_{13} \right)^2} \simeq 0.039$ eV (or $\sqrt{\left(m^{}_2 s^2_{12} c^2_{13}\right)^2 + \left(m^{}_1 c^2_{12} c^2_{13} + m^{}_3 s^2_{13} \right)^2} \simeq 0.052$ eV).

\begin{figure}
\begin{center}
\subfigure{}{
\includegraphics[width=0.45\textwidth]{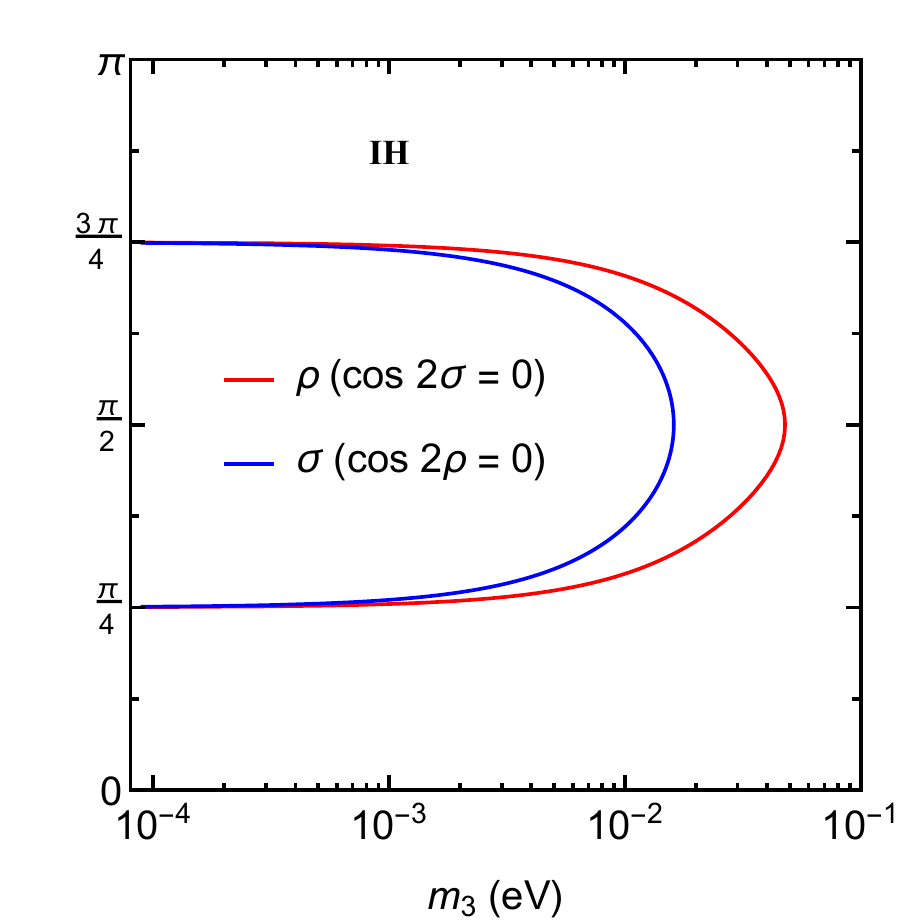}
}
\subfigure{}{
\includegraphics[width=0.45\textwidth]{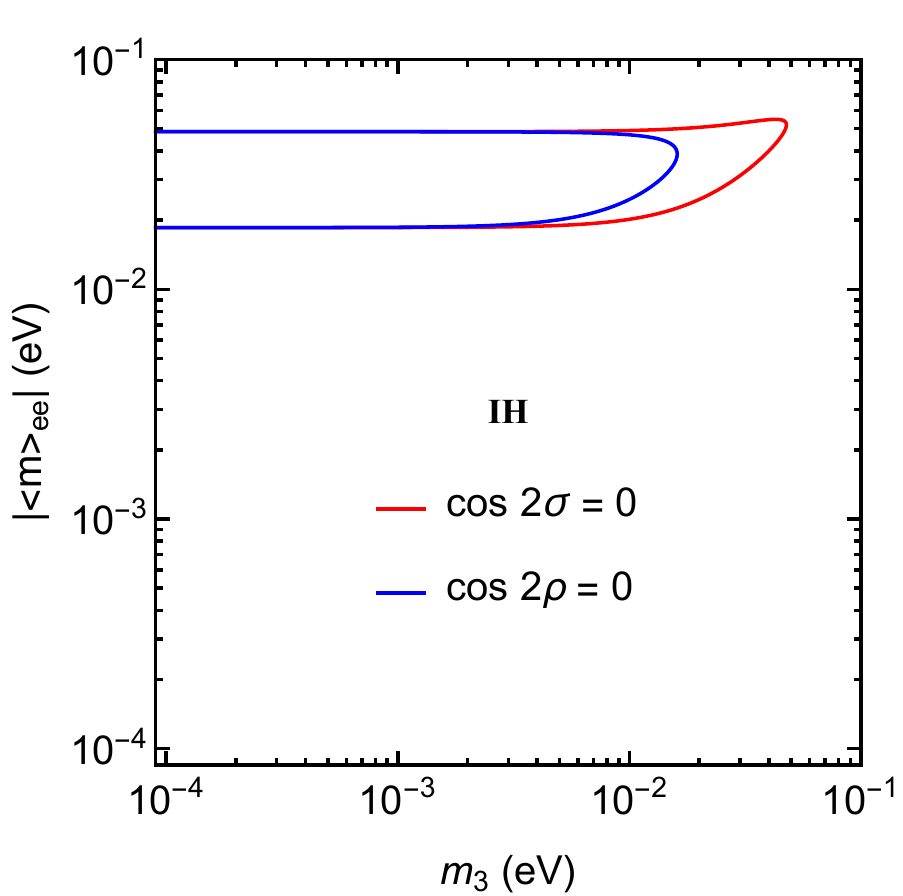}
}
\end{center}
\caption{ Left: $\rho$ (or $\sigma$) as a function of $m^{}_3$ for $\cos 2\sigma = 0$ (or $\cos 2\rho = 0$) in the case of two sides of equation B being vanishing. Right: $\left| \langle m \rangle^{}_{ee} \right|$ as a function of $m^{}_3$ for $\cos 2\sigma = 0$ (or $\cos 2\rho = 0$) in the case of two sides of equation B being vanishing.}
\end{figure}

\subsection{C\&F}

In the case of two sides of equations C and F being vanishing, equations AC, BC, CD and EF become ineffective. We are left with three independent constraint equations (i.e., equation AG and two independent ones of equations AB, AD and BD). But, as discussed at the end of section 2, there are three new constraint equations which are given by Eq. (\ref{28}) and lead to the following relations for $\bar R^{}_{\alpha \beta} $ and $\bar I^{}_{\alpha \beta}$
\begin{eqnarray}
\bar I^{}_{ee} = \bar I^{}_{\mu \mu} + \bar I^{}_{\tau \tau} \;, \hspace{1cm} \bar R^{}_{\mu \mu} = \bar R^{}_{\tau \tau} \;, \hspace{1cm} \bar I^{}_{e \mu} = \bar I^{}_{e \tau}  \;.
\label{42}
\end{eqnarray}
By taking these relations, the expressions for the surviving constraint equations  can be simplified to some extent. In total, the number of independent constraint equations  gets increased by one compared to in the general case. So one neutrino mass sum rule will arise.

With the help of Eq. (\ref{29.1}), one can easily get the desired neutrino mass sum rule as
\begin{eqnarray}
m^{}_1 \sin 2 \rho - m^{}_2 \sin 2 \sigma = 0 \;,
\label{44}
\end{eqnarray}
from any one of the relations in Eq. (\ref{42}). This can be verified by taking the relations in Eq. (\ref{42}) in the expressions for the neutrino masses in combination with the Majorana CP phases in Eq. (\ref{25}).
We discuss the implications of this sum rule in three cases:
(1) In the case of $\sin 2\rho = \sin 2 \sigma = 0$, the neutrino masses become independent of the Majorana CP phases. As a result of ${\rm Im}(m^{}_1 e^{2 {\rm i} \rho}) = {\rm Im}(m^{}_2 e^{2 {\rm i} \sigma}) = 0$, an additional constraint equation
\begin{eqnarray}
\bar I^{}_{\mu \mu} + \bar I^{}_{\tau \tau}  = 2 \bar I^{}_{\mu \tau} \;,
\label{44.1}
\end{eqnarray}
arises from Eq. (\ref{25}) under the condition of Eq. (\ref{42}).
It is found that Eq. (\ref{44.1}) together with Eq. (\ref{42}) will lead us to a situation where two sides of equations A and G are also vanishing. As mentioned in subsection 3.1, this happens to be the case of two sides of equations A, C, F and G being vanishing which will be discussed in subsection 3.6.
(2) In the case of $m^{}_1 = 0$, one immediately obtains $\sigma = 0$ or $\pi/2$.
As a result of ${\rm Re}(m^{}_1 e^{2 {\rm i} \rho}) = {\rm Im}(m^{}_1 e^{2 {\rm i} \rho}) = {\rm Im}(m^{}_2 e^{2 {\rm i} \sigma}) = 0$, two additional constraint equations
arise from Eq. (\ref{25}) under the condition of Eq. (\ref{42}). In this case, $\left| \langle m \rangle^{}_{ee} \right|$
takes a value of $m^{}_2 s^2_{12} c^2_{13}- m^{}_3 s^2_{13} \simeq 0.002$ eV for $\sigma = 0$ or $m^{}_2 s^2_{12} c^2_{13} + m^{}_3 s^2_{13} \simeq 0.004$ eV for $\sigma = \pi/2$.
(3) In the case of $\sin2 \rho$ and $\sin 2\sigma \neq 0$, we present $\sin2\sigma/\sin2\rho$ as a function of the lightest neutrino mass in the NH and IH cases in Fig. 3. In the IH case, it takes a value close to 1 in the whole mass range as a result of $m^{}_1 \simeq m^{}_2$. In the NH case, its value is very small for  vanishingly small $m^{}_1$ but approaches 1 for $m^{}_1 \simeq m^{}_2 \simeq 0.1$ eV.

\begin{figure}
\begin{center}
\includegraphics[width=0.45\textwidth]{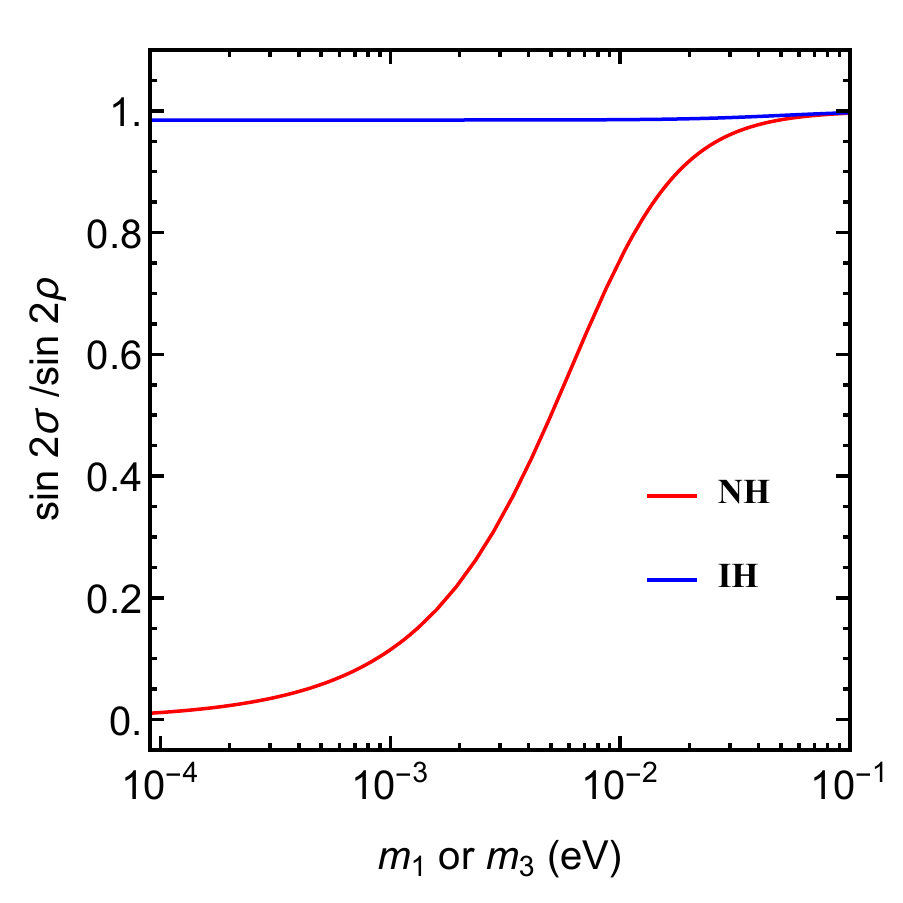}
\end{center}
\caption{$\sin2\sigma/\sin2\rho$ as a function of the lightest neutrino mass ($m^{}_1$ in the NH case or $m^{}_3$ in the IH case) in the case of two sides of equations C and F being vanishing.}
\end{figure}

\subsection{D\&E}

In the case of two sides of equations D and E being vanishing, equations AD, BD, CD and EF become ineffective. We are left with three independent constraint equations (i.e., equation AG and two independent ones of equations AB, AC and BC). But, as discussed at the end of section 2, there are three new constraint equations which are given by Eq. (\ref{29}) and lead to the following relations for $\bar R^{}_{\alpha \beta} $ and $\bar I^{}_{\alpha \beta}$
\begin{eqnarray}
&& \hspace{-0.7cm} \bar R^{}_{ee} + \bar R^{}_{\mu \tau} - \frac{3}{2} \left(\bar R^{}_{\mu \mu} +\bar  R^{}_{\tau \tau}\right)
= -{\rm sgn} \left(\bar I^{}_{e \mu} + \bar I^{}_{e \tau}\right)
\sqrt{2 \left( \bar I^{}_{e \mu} + \bar I^{}_{e \tau}\right)^2 + \left(\bar R^{}_{ee} + \bar R^{}_{\mu \tau} + \frac{ \bar R^{}_{\mu \mu} +\bar  R^{}_{\tau \tau} } {2}\right)^2 }  \;, \nonumber \\
&& \hspace{-0.7cm} \bar I^{}_{\mu \mu} = \bar I^{}_{\tau \tau} \;, \hspace{1cm} \bar R^{}_{e \mu} =\bar R^{}_{e \tau} \;.
\label{45}
\end{eqnarray}
By taking these relations, the expressions for the surviving constraint equations  can be simplified to some extent. In total, the number of independent constraint equations  gets increased by one compared to in the general case. So one neutrino mass sum rule will arise.

With the help of Eq. (\ref{29.1}), one can easily get the desired neutrino mass sum rule as
\begin{eqnarray}
m^{}_1 \cos 2 \rho - m^{}_2 \cos 2 \sigma = 0 \;,
\label{47}
\end{eqnarray}
from any one of the relations in Eq. (\ref{45}). This can be verified by taking the relations in Eq. (\ref{45}) in the expressions for the neutrino masses in combination with the Majorana CP phases in Eq. (\ref{25}).
We discuss the implications of this sum rule in three cases:
(1) In the case of $\cos 2\rho = \cos 2\sigma = 0$, the neutrino masses become independent of the Majorana CP phases. As a result of ${\rm Re}(m^{}_1 e^{2 {\rm i} \rho}) = {\rm Re}(m^{}_2 e^{2 {\rm i} \sigma}) = 0$, an additional constraint equation
\begin{eqnarray}
\bar R^{}_{\mu \mu} + \bar R^{}_{\tau \tau} = 2 \bar R^{}_{\mu \tau} \;,
\label{48}
\end{eqnarray}
arises from Eq. (\ref{25}) under the condition of Eq. (\ref{45}). For the possible combinations $[\rho, \sigma] = [\pi/4, \pi/4], [\pi/4, 3\pi/4], [3\pi/4, \pi/4]$ and $[3\pi/4, 3\pi/4]$, we present $\left| \langle m \rangle^{}_{ee} \right|$ as a function of the lightest neutrino mass in the NH and IH cases in Fig. 4: $\left| \langle m \rangle^{}_{ee} \right| = \sqrt{ \left(m^{}_1 c^2_{12} + m^{}_2 s^2_{12} \right)^2 c^4_{13} + m^2_3 s^4_{13}}$ for $\rho = \sigma$ or $\sqrt{ \left(m^{}_1 c^2_{12} - m^{}_2 s^2_{12}\right)^2 c^4_{13} + m^2_3 s^4_{13}}$ for $\rho \neq \sigma$. (2) In the case of $m^{}_1 = 0$, one immediately obtains $\sigma = \pi/4$ or $3\pi/4$.
As a result of ${\rm Re}(m^{}_1 e^{2 {\rm i} \rho}) = {\rm Im}(m^{}_1 e^{2 {\rm i} \rho}) = {\rm Re}(m^{}_2 e^{2 {\rm i} \sigma}) = 0$, two additional constraint equations
arise from Eq. (\ref{25}) under the condition of Eq. (\ref{45}). In this case, $\left| \langle m \rangle^{}_{ee} \right|$
takes a value of $\sqrt{m^{2}_2 s^4_{12} c^4_{13} + m^{2}_3 s^4_{13} } \simeq 0.003$ eV. (3) In the case of $\cos 2\rho$ and $\cos 2 \sigma \neq 0$, one can also present $\cos 2\sigma/\cos 2\rho$ as a function of the lightest neutrino mass. The result is the same as that for $\sin 2\sigma/ \sin 2\rho$ in subsection 3.3.

\begin{figure}
\begin{center}
\subfigure{}{
\includegraphics[width=0.45\textwidth]{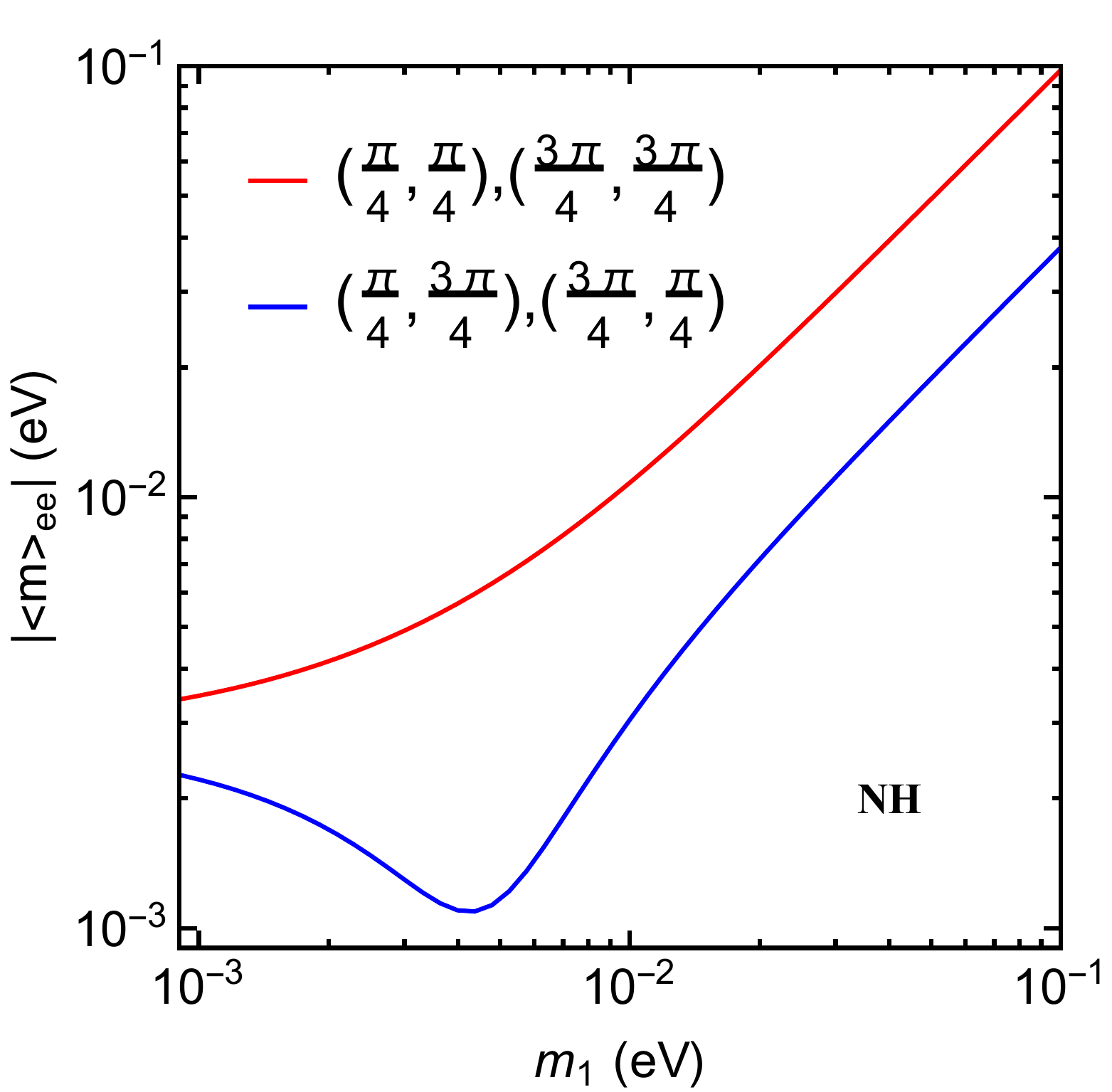}
}
\subfigure{}{
\includegraphics[width=0.45\textwidth]{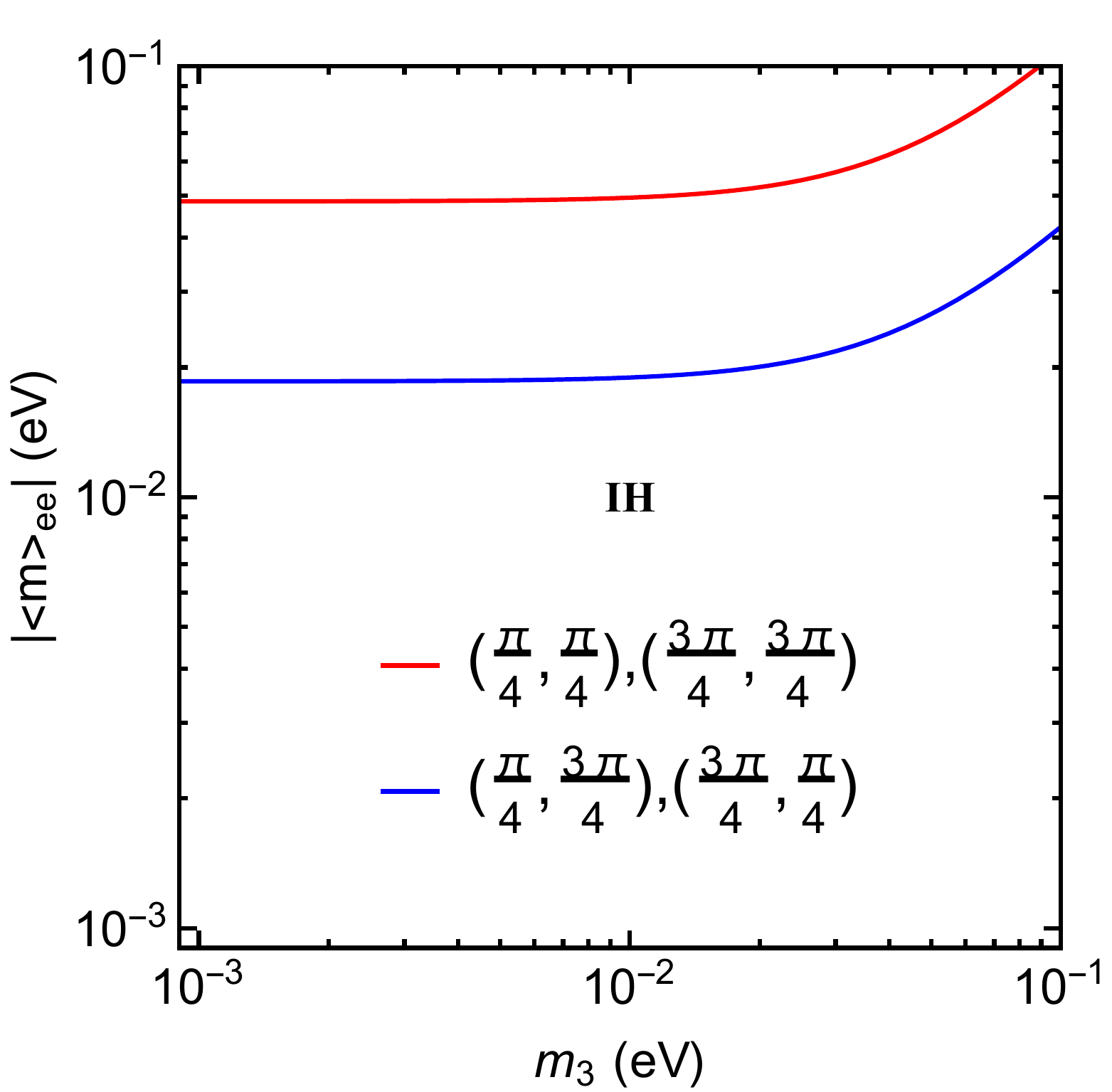}
}
\end{center}
\caption{$\left| \langle m \rangle^{}_{ee} \right|$ as a function of the lightest neutrino mass ($m^{}_1$ in the NH case or $m^{}_3$ in the IH case) for the possible combinations $[\rho, \sigma] = [\pi/4, \pi/4], [\pi/4, 3\pi/4], [3\pi/4, \pi/4]$ and $[3\pi/4, 3\pi/4]$ in the case of two sides of equation D being vanishing.}
\end{figure}

\subsection{A\&B\&G}

In the case of two sides of equations A, B and G being vanishing, equations AB, AC, AD, AG, BC and BD become ineffective. We are left with two constraint equations (i.e., equations CD and EF). But there are five new constraint equations which are given by Eqs. (\ref{27}, \ref{34}) and lead to the following relations for $\bar R^{}_{\alpha \beta} $ and $\bar I^{}_{\alpha \beta}$ (see Eqs. (\ref{30}, \ref{35}))
\begin{eqnarray}
&& \bar R^{}_{e \mu} = -\bar R^{}_{e \tau} \;, \hspace{1cm} \bar I^{}_{ee} = 0 \;, \hspace{1cm} -2 \bar I^{}_{\mu \tau} = \bar I^{}_{\mu \mu} + \bar I^{}_{\tau \tau} \;, \nonumber \\
&& \bar I^{}_{e \mu} = - \bar I^{}_{e \tau} \;, \hspace{1cm} -2 \left( \bar R^{}_{ee} + \bar R^{}_{\mu \tau} \right) =  \bar R^{}_{\mu \mu} + \bar R^{}_{\tau \tau} \;.
\label{48.1}
\end{eqnarray}
It is easy to see that these relations can be recombined into
\begin{eqnarray}
&& \bar M^{}_{e \mu} = -\bar M^{}_{e \tau} \;,  \hspace{1cm}  -2 \left( \bar M^{}_{ee} + \bar M^{}_{\mu \tau} \right) =  \bar M^{}_{\mu \mu} + \bar M^{}_{\tau \tau} \;, \hspace{1cm} \bar M^{}_{ee} \ {\rm being \ real} \;.
\label{48.2}
\end{eqnarray}
By taking these relations, the expressions for the surviving constraint equations  can be simplified to some extent. In total, the number of independent constraint equations  gets increased by two compared to in the general case. So two neutrino mass sum rules will arise.

It turns out that the desired neutrino mass sum rules are the same as those in Eqs. (\ref{32}, \ref{37}).
This can be verified by taking the relations in Eq. (\ref{48.1}) in the expressions for the neutrino masses in combination with the Majorana CP phases in Eq. (\ref{25}).
As discussed in subsection 3.2, these sum rules can only be fulfilled in the IH case. In Fig. 5, we present $\rho$, $\sigma$ and $\left| \langle m \rangle^{}_{ee} \right|$ as functions of the lightest neutrino mass $m^{}_3$. There is a lower value $0.021$ eV of $m^{}_3$ at which $\rho$ and $\sigma$ respectively take the values $\pi/2$ and 0. As discussed in subsection 3.1, $\sin 2\rho = \sin 2\sigma = 0$ (which gives an additional constraint equation given by Eq. (\ref{33})) will lead us to a situation where two sides of equations C and F are also vanishing. This happens to be the case of two sides of equations A, B, C, F and G being vanishing which will be discussed in subsection 3.10. On the other hand, $\left| \langle m \rangle^{}_{ee} \right|$ is found to be equal to $m^{}_3$: The sum rules in Eqs. (\ref{32}, \ref{37}) can be reorganized into a single complex equation
\begin{eqnarray}
m^{}_1 c^2_{12} e^{2{\rm i}\rho} + m^{}_2 s^2_{12} e^{2{\rm i}\sigma} + m^{}_3 = 0 \;,
\label{49}
\end{eqnarray}
which immediately leads us to $\left| \langle m \rangle^{}_{ee} \right| = m^{}_3$.

\begin{figure}
\begin{center}
\subfigure{}{
\includegraphics[width=0.45\textwidth]{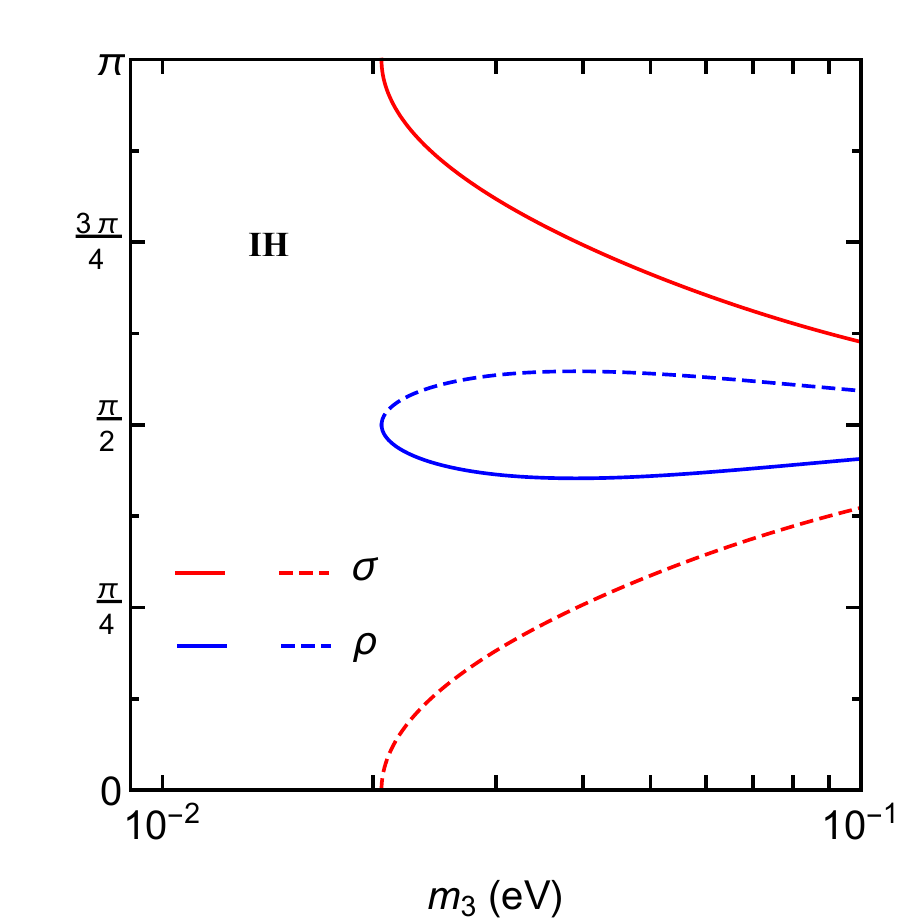}
}
\subfigure{}{
\includegraphics[width=0.46\textwidth]{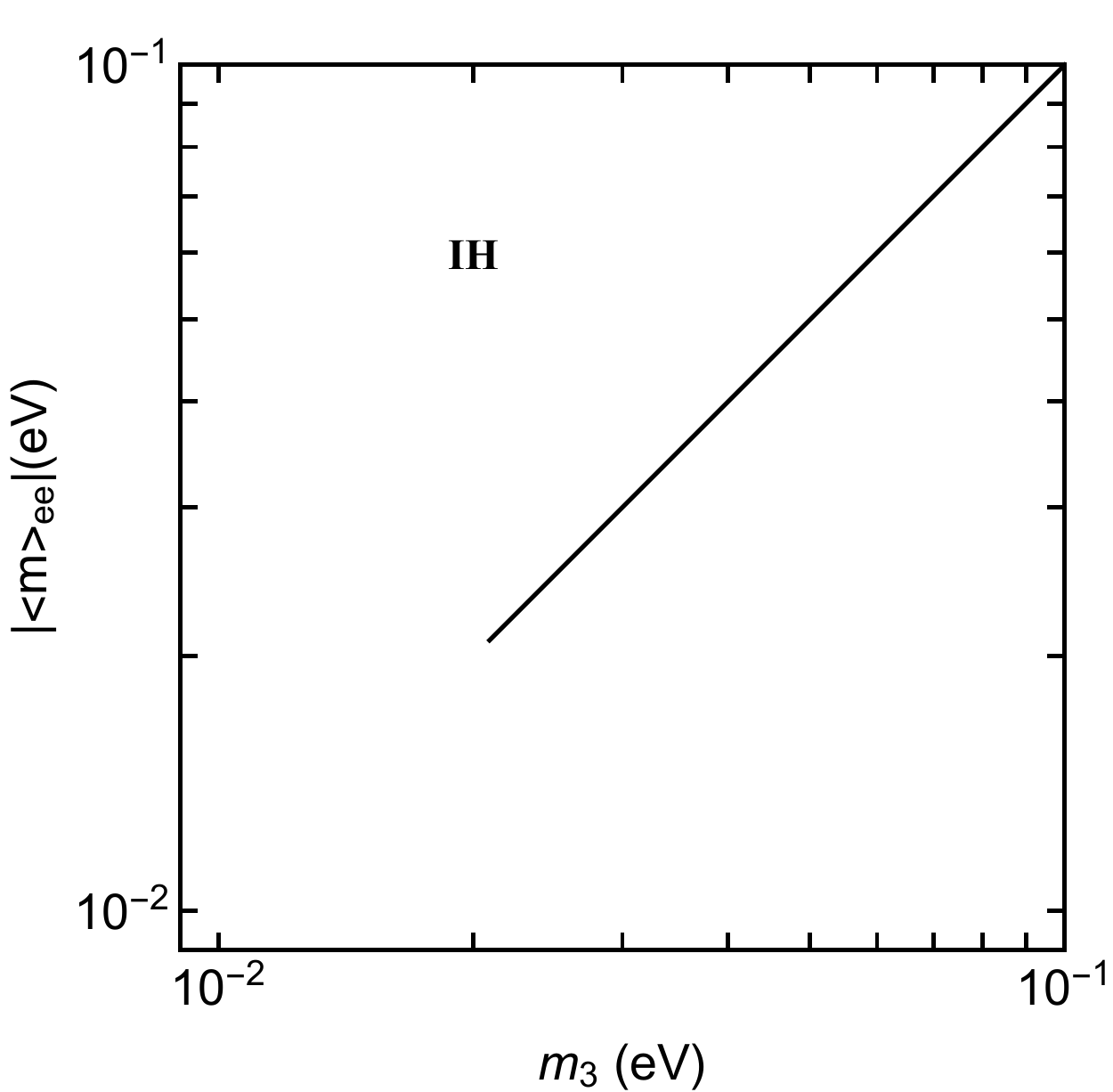}
}
\end{center}
\caption{$\rho$, $\sigma$ and $\left| \langle m \rangle^{}_{ee} \right|$ as functions of the lightest neutrino mass $m^{}_3$ in the case of two sides of equations A, B and G being vanishing. }
\end{figure}

\subsection{A\&C\&F\&G}

In the case of two sides of equations A, C, F and G being vanishing, equations AB, AC, AD, AG, BC, CD and EF become ineffective. We are left with only one constraint equation (i.e., equation BD). But there are six new constraint equations which are given by Eqs. (\ref{27}, \ref{28}) and lead to the following relations for $\bar R^{}_{\alpha \beta} $ and $\bar I^{}_{\alpha \beta}$ (see Eqs. (\ref{30}, \ref{42}))
\begin{eqnarray}
&& \bar R^{}_{e \mu} = -\bar R^{}_{e \tau} \;,  \hspace{1cm} \bar I^{}_{e \mu} = \bar I^{}_{e \tau} \;, \hspace{1cm} \bar R^{}_{\mu \mu} = \bar R^{}_{\tau \tau} \;, \hspace{1cm} \bar I^{}_{\mu \mu} = - \bar I^{}_{\tau \tau} \;, \hspace{1cm} \bar I^{}_{ee} = \bar I^{}_{\mu \tau} = 0 \;.
\label{49.1}
\end{eqnarray}
It is easy to see that these relations can be recombined into
\begin{eqnarray}
\bar M^{}_{e\mu} = - \bar M^*_{e\tau} \;, \hspace{1cm} \bar M^{}_{\mu\mu} = \bar M^*_{\tau\tau}  \;, \hspace{1cm}
\bar M^{}_{ee} \ {\rm and} \ \bar M^{}_{\mu\tau} \ {\rm being \ real}  \;.
\label{50}
\end{eqnarray}
By taking these relations, the expression for the surviving constraint equation  can be simplified to some extent. It is interesting to find that the texture thus obtained can reproduce the specific texture of $M^{}_\nu$ given by the $\mu$-$\tau$ reflection symmetry:
In view of the definition $\bar M^{}_{\alpha \beta} = M^{}_{\alpha \beta} e^{-{\rm i}(\phi^{}_\alpha+ \phi^{}_\beta)}$, the relations in Eq. (\ref{50}) will become those in Eq. (\ref{8}) by taking $\phi^{}_e =\pi/2$ and $\phi^{}_\mu = - \phi^{}_\tau$. In total, the number of independent constraint equations  gets increased by two compared to in the general case. So two neutrino mass sum rules will arise.

It turns out that the desired neutrino mass sum rules are the same as those in Eqs. (\ref{32}, \ref{44}), implying that the Majorana CP phases take trivial values (i.e., $\rho, \sigma = 0$ or $\pi/2$). This can be verified by taking the relations in Eq. (\ref{49.1}) in the expressions for the neutrino masses in combination with the Majorana CP phases in Eq. (\ref{25}). For the possible combinations $[\rho, \sigma] = [0, 0], [0, \pi/2], [\pi/2, 0]$ and $[\pi/2, \pi/2]$, we present $\left| \langle m \rangle^{}_{ee} \right|$ as a function of the lightest neutrino mass in the NH and IH cases in Fig. 6. In the NH case, three terms of $\left| \langle m \rangle^{}_{ee} \right|$ add constructively to a maximal level for $[\rho, \sigma] = [\pi/2, \pi/2]$ but will cancel out (i.e., $\left| \langle m \rangle^{}_{ee} \right| = 0$) at $m^{}_1 = 0.002$ eV (or 0.007 eV) for $[\rho, \sigma] = [\pi/2, 0]$ (or $[0, \pi/2]$). In the IH case,
$\left| \langle m \rangle^{}_{ee} \right|$ is dominated by the first two terms since the third one is highly suppressed. For $\rho = \sigma$ (or $\rho \neq \sigma$), $\left| \langle m \rangle^{}_{ee} \right|$ approximates to $m^{}_1 c^2_{12} c^2_{13} + m^{}_2 s^2_{12} c^2_{13} $ (or $m^{}_1 c^2_{12} c^2_{13} -  m^{}_2 s^2_{12} c^2_{13}$) .

\begin{figure}
\begin{center}
\subfigure{}{
\includegraphics[width=0.45\textwidth]{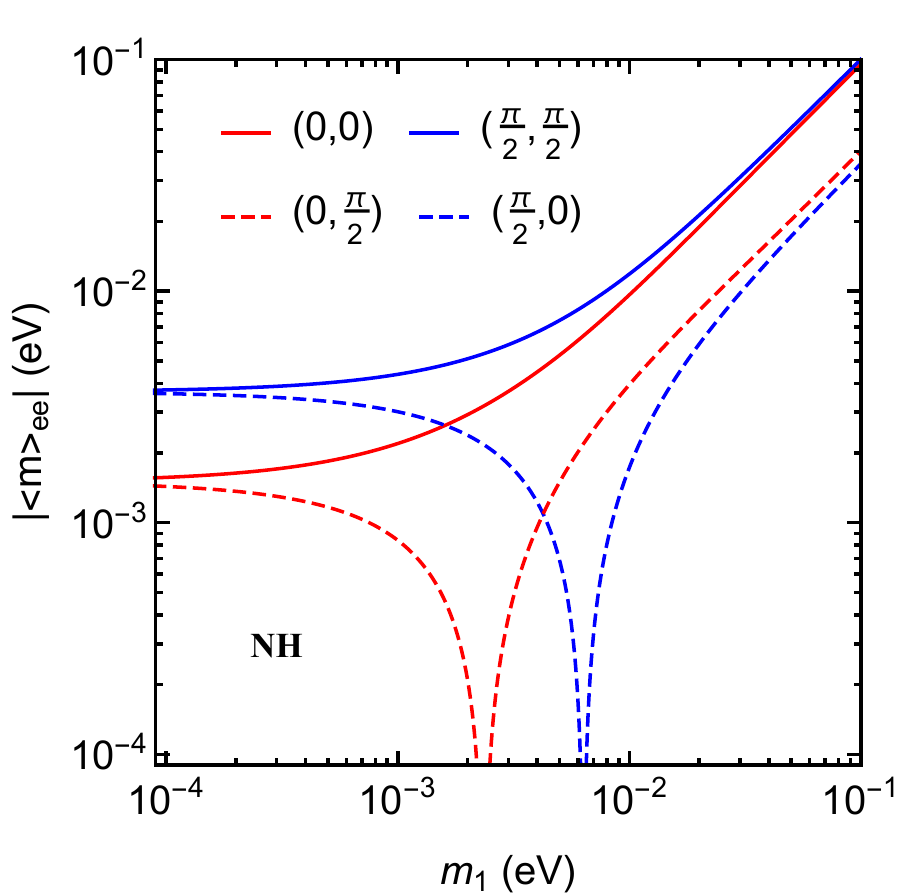}
}
\subfigure{}{
\includegraphics[width=0.45\textwidth]{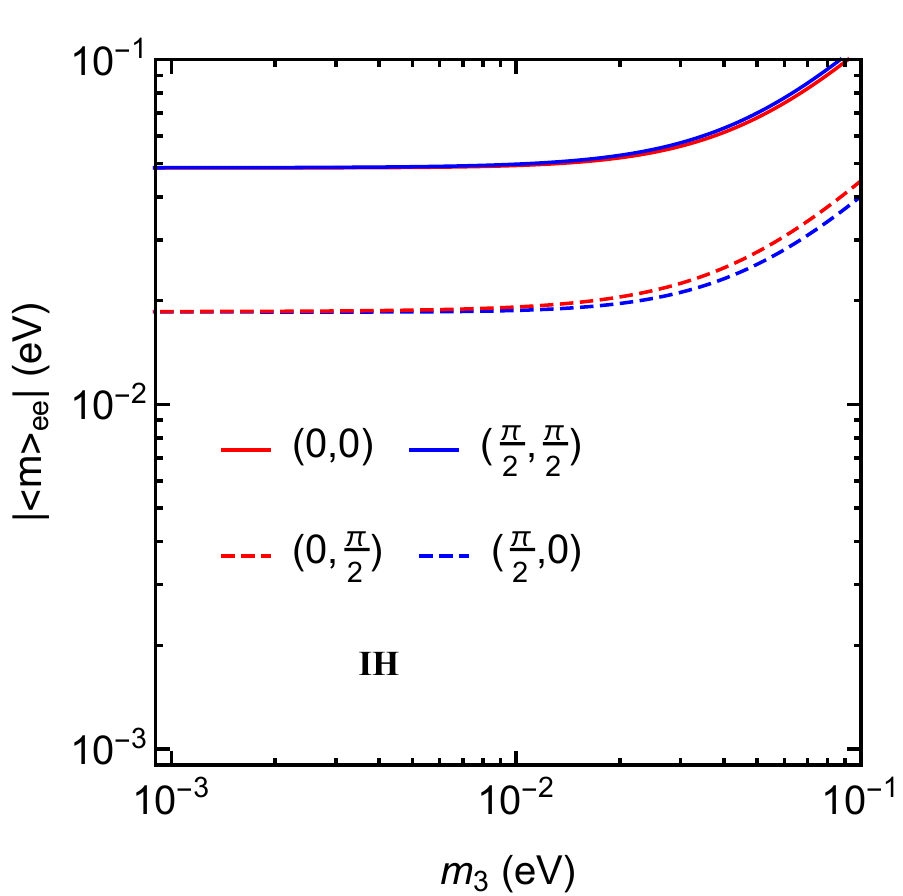}
}
\end{center}
\caption{$\left| \langle m \rangle^{}_{ee} \right|$ as a function of the lightest neutrino mass ($m^{}_1$ in the NH case or $m^{}_3$ in the IH case) for the possible combinations $[\rho, \sigma] = [0, 0], [0, \pi/2], [\pi/2, 0]$ and $[\pi/2, \pi/2]$ in the case of two sides of equations A, C, F and G being vanishing.}
\end{figure}

\subsection{A\&D\&E\&G}

In the case of two sides of equations A, D, E and G being vanishing, equations AB, AC, AD, AG, BD, CD and EF become ineffective. We are left with only one constraint equation (i.e., equation BC). But there are six new constraint equations which are given by Eqs. (\ref{27}, \ref{29}) and lead to the following relations for $\bar R^{}_{\alpha \beta} $ and $\bar I^{}_{\alpha \beta}$ (see Eqs. (\ref{30}, \ref{45}))
\begin{eqnarray}
&& \bar R^{}_{ee} + \bar R^{}_{\mu \tau} - \frac{3}{2} \left(\bar R^{}_{\mu \mu} +\bar  R^{}_{\tau \tau}\right)
= -{\rm sgn} \left(\bar I^{}_{e \mu} + \bar I^{}_{e \tau}\right)
\sqrt{2 \left( \bar I^{}_{e \mu} + \bar I^{}_{e \tau}\right)^2 + \left(\bar R^{}_{ee} + \bar R^{}_{\mu \tau} + \frac{ \bar R^{}_{\mu \mu} +\bar  R^{}_{\tau \tau} } {2}\right)^2 }  \;, \nonumber \\
&& \bar R^{}_{e \mu} = \bar R^{}_{e \tau} = \bar I^{}_{ee} = 0 \;, \hspace{1cm}   \bar I^{}_{\mu \mu}  = \bar I^{}_{\tau \tau} = - \bar I^{}_{\mu \tau} \;.
\label{50.1}
\end{eqnarray}
In total, the number of independent constraint equations  gets increased by two compared to in the general case. So two neutrino mass sum rules will arise.

It turns out that the desired neutrino mass sum rules are the same as those in Eqs. (\ref{32}, \ref{47}).
This can be verified by taking the relations in Eq. (\ref{50.1}) in the expressions for the neutrino masses in combination with the Majorana CP phases in Eq. (\ref{25}). In Figs. 7-8, we present $\rho$, $\sigma$ and $\left| \langle m \rangle^{}_{ee} \right|$ as functions of the lightest neutrino mass in the NH and IH cases. In the NH case, there is a lower value 0.004 eV of $m^{}_1$ at which $\rho$ and $\sigma$ respectively take the values $\pi/4$ and $3\pi/4$ (or $3\pi/4$ and $\pi/4$). As discussed in subsection 3.4, $\cos 2\rho = \cos 2\sigma = 0$ gives an additional constraint equation given by Eq. (\ref{48}). At this lower value of $m^{}_1$, $\left| \langle m \rangle^{}_{ee} \right|$ takes a value of $m^{}_3 s^2_{13}$. For $m^{}_1 \simeq m^{}_2 \simeq 0.1$ eV, Eqs. (\ref{32}, \ref{47}) give
\begin{eqnarray}
\frac {\sin 2 \sigma}{\sin 2\rho} \simeq - \frac{c^2_{12}}{s^2_{12}} \;, \hspace{1cm}\
\cos 2\rho \simeq \cos 2\sigma \;.
\label{51}
\end{eqnarray}
Only for $\rho \simeq \pi- \sigma \simeq 0$ or $\pi/2$, can these two relations be fulfilled simultaneously. Consequently, $\left| \langle m \rangle^{}_{ee} \right|$ approximates to $m^{}_1 c^2_{12} c^2_{13} + m^{}_2 s^2_{12} c^2_{13} \pm m^{}_3 s^2_{13}$ in this mass range. In the IH case, we have $m^{}_1 \simeq m^{}_2$ and thus $\rho \simeq \pi- \sigma \simeq 0$ or $\pi/2$ and $\left| \langle m \rangle^{}_{ee} \right| \simeq m^{}_1 c^2_{12} c^2_{13} + m^{}_2 s^2_{12} c^2_{13} \pm m^{}_3 s^2_{13}$ in the whole mass range.

\begin{figure}
\begin{center}
\subfigure{}{
\includegraphics[width=0.45\textwidth]{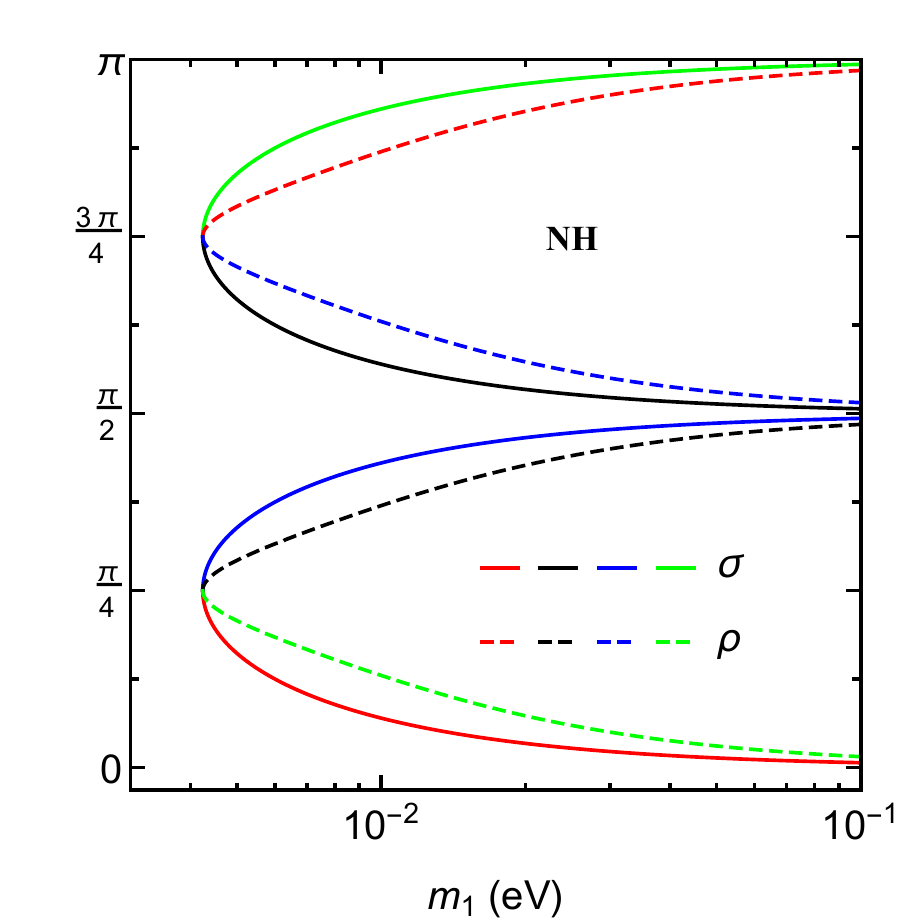}
}
\subfigure{}{
\includegraphics[width=0.45\textwidth]{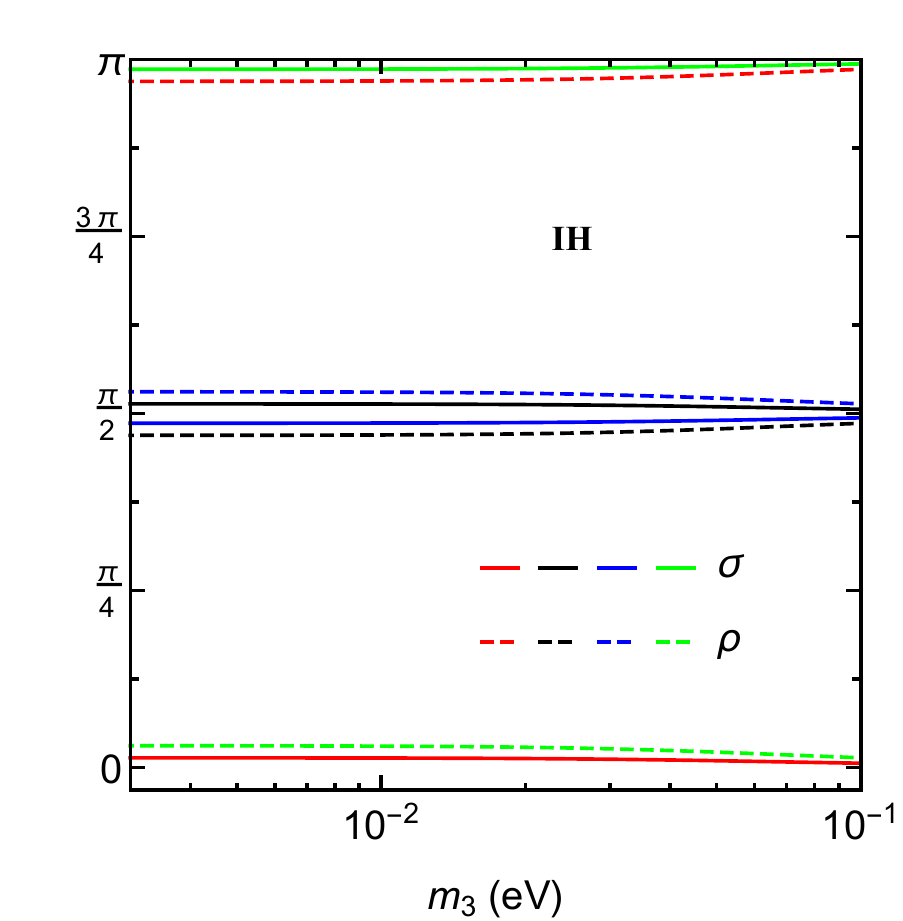}
}
\end{center}
\caption{$\rho$ and $\sigma$ as functions of the lightest neutrino mass ($m^{}_1$ in the NH case or $m^{}_3$ in the IH case) in the case of two sides of equations A, D, E and G being vanishing.}
\end{figure}

\begin{figure}
\begin{center}
\includegraphics[width=0.45\textwidth]{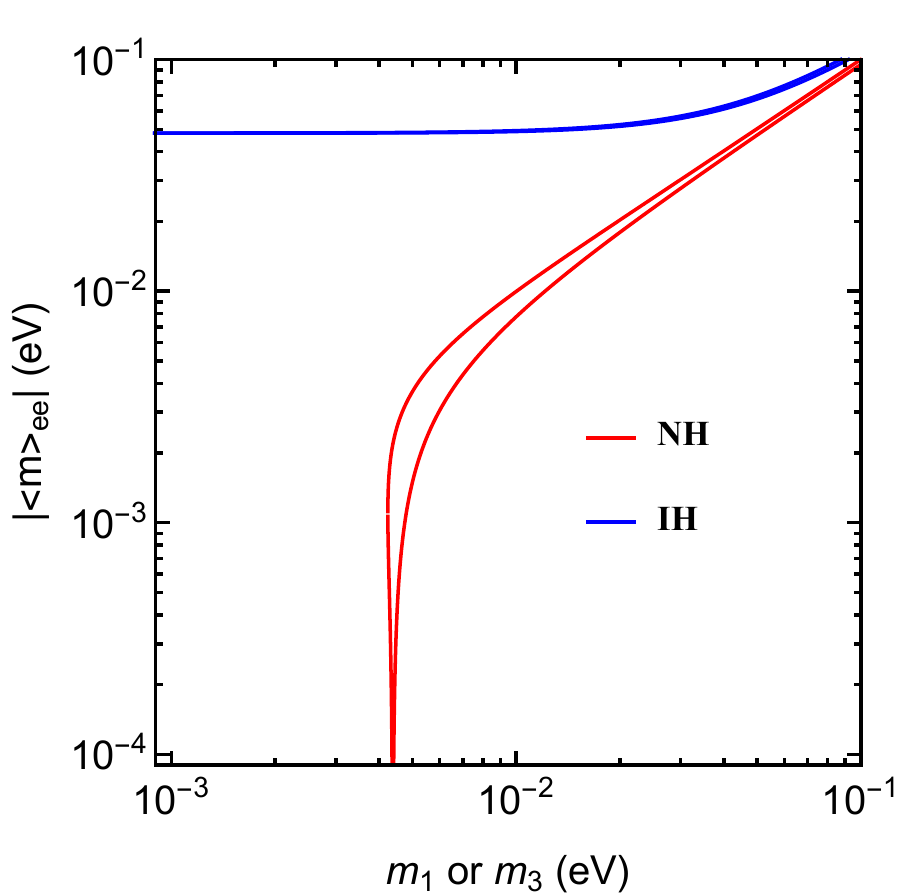}
\end{center}
\caption{$\left| \langle m \rangle^{}_{ee} \right|$ as a function of the lightest neutrino mass ($m^{}_1$ in the NH case or $m^{}_3$ in the IH case) in the case of two sides of equations A, D, E and G being vanishing.}
\end{figure}

\subsection{B\&C\&F}

In the case of two sides of equations B, C and F being vanishing, equations AB, AC, BC, BD, CD and EF become ineffective. We are left with two constraint equations (i.e., equations AD and AG). But there are five new constraint equations which are given by Eqs. (\ref{34}, \ref{28}) and lead to the following relations for $\bar R^{}_{\alpha \beta} $ and $\bar I^{}_{\alpha \beta}$ (see Eqs. (\ref{35}, \ref{42}))
\begin{eqnarray}
\bar I^{}_{ee} = \bar I^{}_{\mu \mu} + \bar I^{}_{\tau \tau} \;, \hspace{1cm} \bar R^{}_{\mu \mu} = \bar R^{}_{\tau \tau} = - \left( \bar R^{}_{ee} + \bar R^{}_{\mu \tau} \right) \;, \hspace{1cm} \bar I^{}_{e \mu} = \bar I^{}_{e \tau} = 0  \;.
\label{51.1}
\end{eqnarray}
By taking these relations, the expressions for the surviving constraint equations  can be simplified to some extent.
In total, the number of independent constraint equations  gets increased by two compared to in the general case. So two neutrino mass sum rules will arise.

It turns out that the desired neutrino mass sum rules are the same as those in Eqs. (\ref{37}, \ref{44}).
This can be verified by taking the relations in Eq. (\ref{51.1}) in the expressions for the neutrino masses in combination with the Majorana CP phases in Eq. (\ref{25}). As discussed in subsection 3.2, these sum rules can only be fulfilled in the IH case. In Fig. 9, we present $\rho$, $\sigma$ and $\left| \langle m \rangle^{}_{ee} \right|$ as functions of the lightest neutrino mass $m^{}_3$. As a result of $m^{}_1 \simeq m^{}_2$ in the IH case, one gets $\rho \simeq \sigma$ or $\pi/2 -\sigma$ from Eq. (\ref{44}). Eq. (\ref{37}) further leads us to $\rho \simeq \sigma \simeq \pi/4$ or $3\pi/4$ for vanishingly small $m^{}_3$ and $\rho \simeq \sigma \simeq \pi/2$ for $m^{}_3 \simeq m^{}_1 \simeq m^{}_2 \simeq 0.1$ eV. Consequently, $\left| \langle m \rangle^{}_{ee} \right|$ approximates to $m^{}_1 c^2_{12} c^2_{13} + m^{}_2 s^2_{12} c^2_{13}$ for these two mass ranges. For the particular value of $m^{}_3 = 0.021$ eV, one of the allowed solutions to Eqs. (\ref{37}, \ref{44}) is $\rho = \pi/2$ and $\sigma = 0$ in which case $\left| \langle m \rangle^{}_{ee} \right|$ becomes $m^{}_1 c^2_{12} c^2_{13} - m^{}_2 s^2_{12}  c^2_{13} + m^{}_3 s^2_{13} \simeq 0.020$ eV. As discussed in subsection 3.3, $\sin 2\rho = \sin 2\sigma = 0$ (which gives an additional constraint equation given by Eq. (\ref{44.1})) will lead us to a situation where two sides of equations A and G are also vanishing. This happens to be the case of two sides of equations A, B, C, F and G being vanishing which will be discussed in subsection 3.10.

\begin{figure}
\begin{center}
\subfigure{}{
\includegraphics[width=0.45\textwidth]{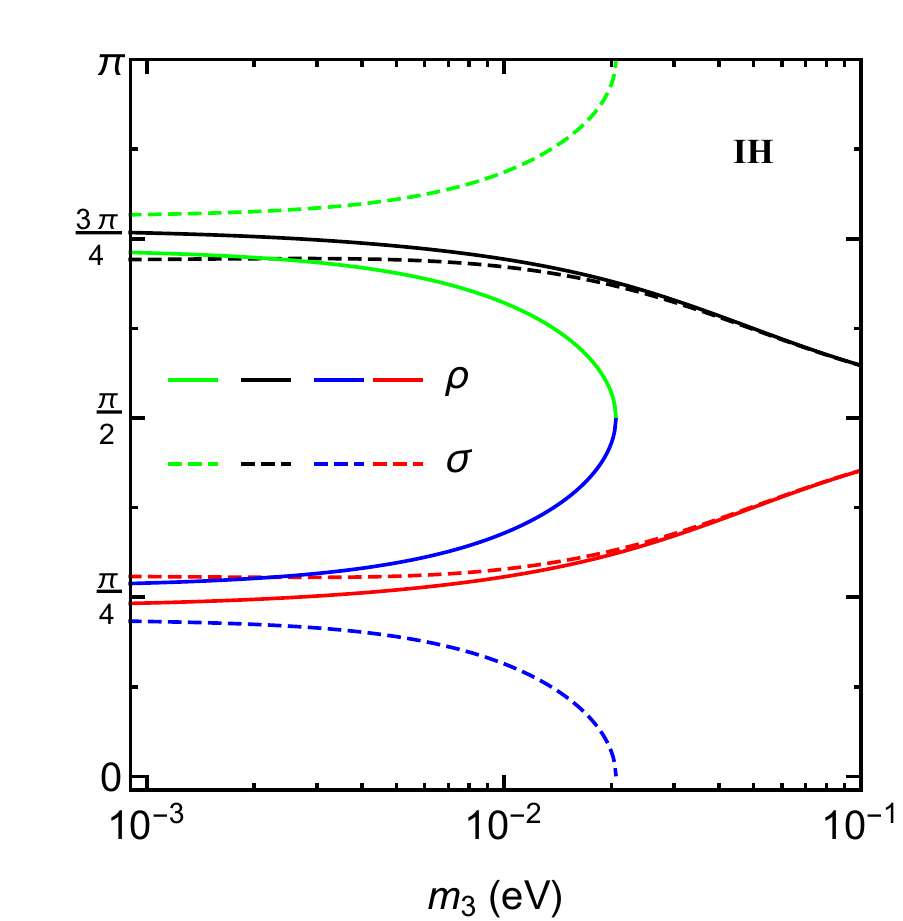}
}
\subfigure{}{
\includegraphics[width=0.46\textwidth]{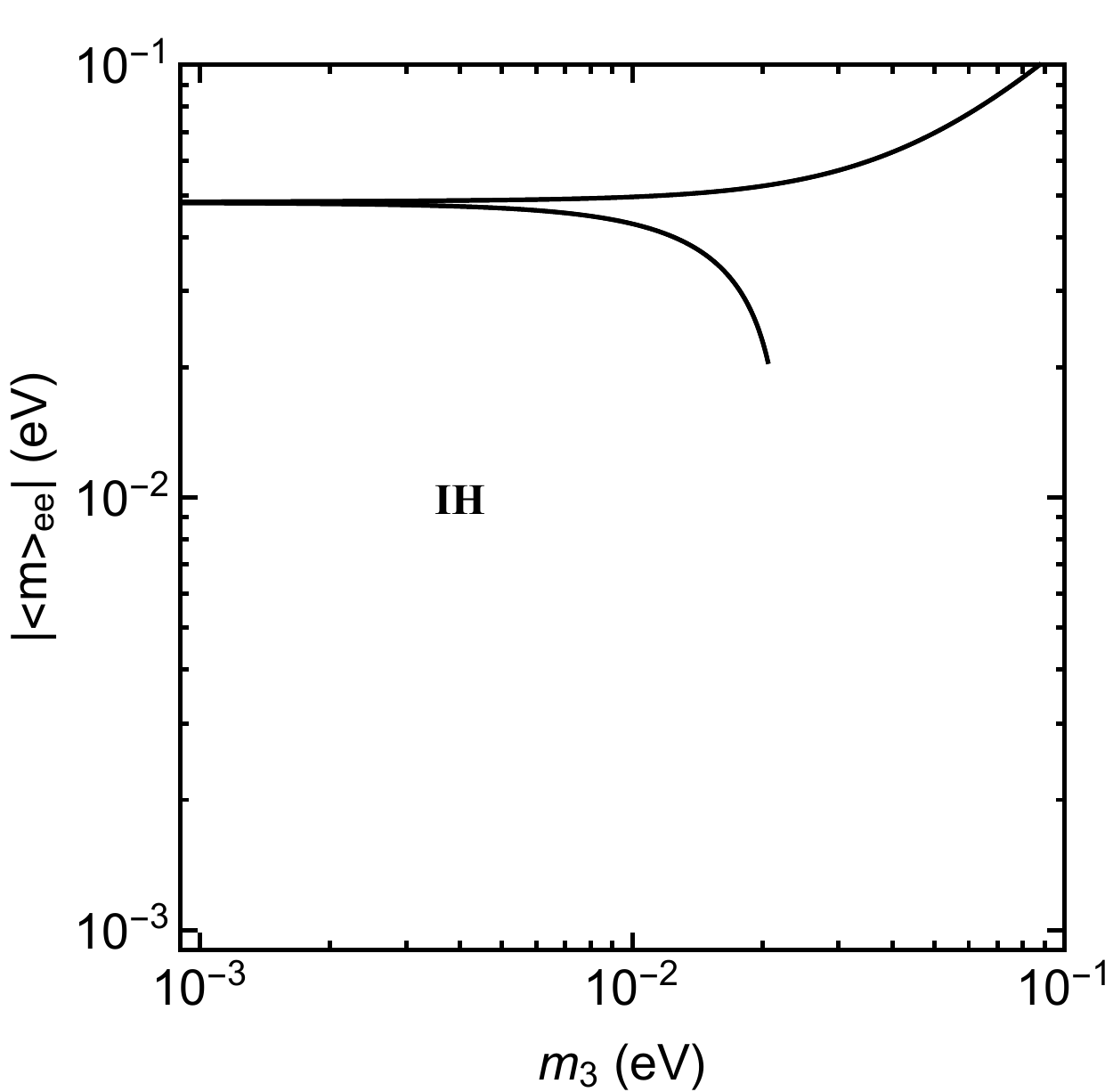}
}
\end{center}
\caption{$\rho$, $\sigma$ and $\left| \langle m \rangle^{}_{ee} \right|$ as functions of the lightest neutrino mass $m^{}_3$ in the case of two sides of equations B, C and F being vanishing.}
\end{figure}

\subsection{B\&D\&E}

In the case of two sides of equations B, D and E being vanishing, equations AB, AD, BC, BD,  CD and EF become ineffective. We are left with two constraint equations (i.e., equations AC and AG). But there are five new constraint equations which are given by Eqs. (\ref{34}, \ref{29}) and lead to the following relations for $\bar R^{}_{\alpha \beta} $ and $\bar I^{}_{\alpha \beta}$ (see Eqs. (\ref{35}, \ref{45}))
\begin{eqnarray}
\bar R^{}_{e \mu} =\bar R^{}_{e \tau} \;, \hspace{1cm} \bar I^{}_{e \mu} = - \bar I^{}_{e \tau} \;, \hspace{1cm} \bar R^{}_{\mu \mu} = - \bar R^{}_{\tau \tau} \;, \hspace{1cm} \bar I^{}_{\mu \mu} = \bar I^{}_{\tau \tau} \;, \hspace{1cm}  \bar R^{}_{ee} = - \bar R^{}_{\mu \tau}  \;.
\label{51.2}
\end{eqnarray}
It is easy to see that these relations can be recombined into
\begin{eqnarray}
\bar M^{}_{e\mu} = \bar M^*_{e\tau} \;, \hspace{1cm} \bar M^{}_{\mu\mu} = - \bar M^*_{\tau\tau}  \;, \hspace{1cm} \bar R^{}_{ee} = - \bar R^{}_{\mu \tau} \;.
\label{51.3}
\end{eqnarray}
By taking these relations, the expressions for the surviving constraint equations can be simplified to some extent.
In total, the number of independent constraint equations  gets increased by two compared to in the general case. So two neutrino mass sum rules will arise.

It turns out that the desired neutrino mass sum rules are the same as those in Eqs. (\ref{37}, \ref{47}).
This can be verified by taking the relations in Eq. (\ref{51.2}) in the expressions for the neutrino masses in combination with the Majorana CP phases in Eq. (\ref{25}).
A combination of these sum rules yields
\begin{eqnarray}
m^{}_1 \cos 2 \rho = m^{}_2 \cos 2\sigma = -m^{}_3 \;.
\label{52}
\end{eqnarray}
Apparently, these relations can only be fulfilled in the IH case. In Fig. 10, we present $\rho$, $\sigma$ and $\left| \langle m \rangle^{}_{ee} \right|$ as functions of the lightest neutrino mass $m^{}_3$. In consideration of $m^{}_1 \simeq m^{}_2$ in the IH case, the results for $\rho$ and $\sigma$ are presented by the same lines. One has $\rho, \sigma \simeq \pi/4$ or $3\pi/4$ for vanishingly small $m^{}_3$ and $\rho \simeq \sigma \simeq \pi/2$ for $m^{}_3 \simeq m^{}_1 \simeq m^{}_2 \simeq 0.1$ eV. Consequently, $\left| \langle m \rangle^{}_{ee} \right|$ takes a value close to $m^{}_1 c^2_{12} c^2_{13} + m^{}_2 s^2_{12} c^2_{13} \simeq 0.049$ eV in the case of $\rho \simeq \sigma $ (or $m^{}_1 c^2_{12} c^2_{13} - m^{}_2 s^2_{12} c^2_{13} \simeq 0.019$ eV in the case of $\rho \simeq \pi/2+ \sigma$) for vanishingly small $m^{}_3$ and $m^{}_1 c^2_{12} c^2_{13} + m^{}_2 s^2_{12} c^2_{13}+ m^{}_3 s^2_{13}$ for $m^{}_3 \simeq m^{}_1 \simeq m^{}_2 \simeq 0.1$ eV.

\begin{figure}
\begin{center}
\subfigure{}{
\includegraphics[width=0.45\textwidth]{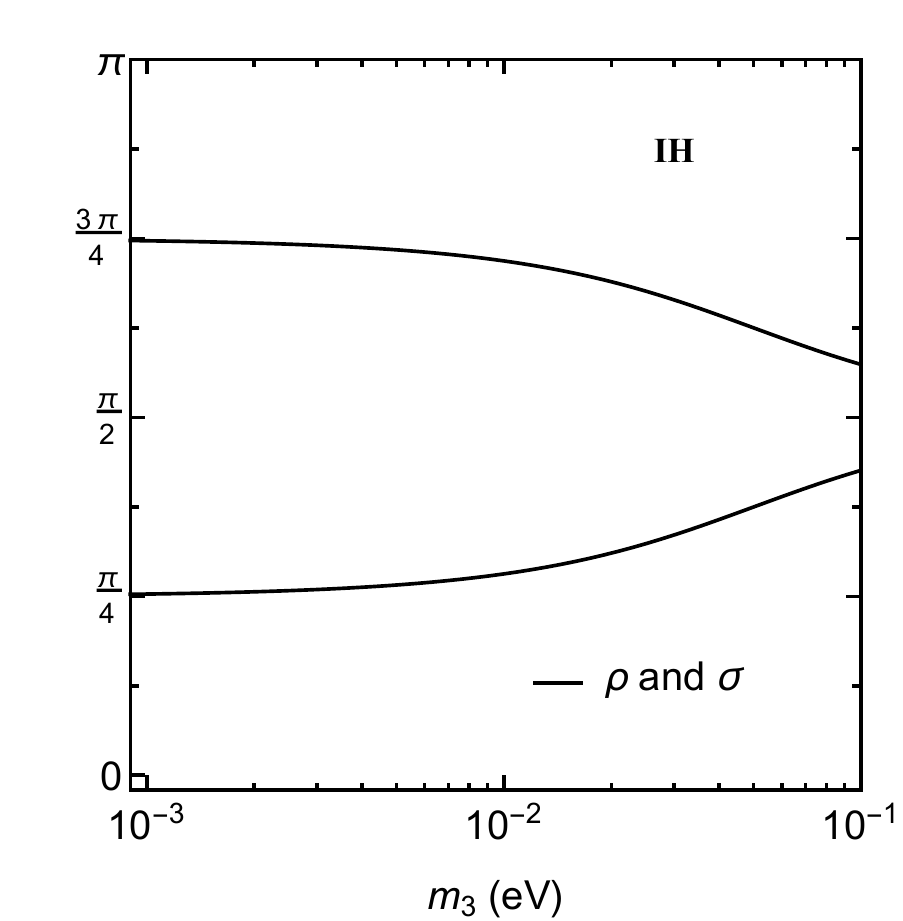}
}
\subfigure{}{
\includegraphics[width=0.46\textwidth]{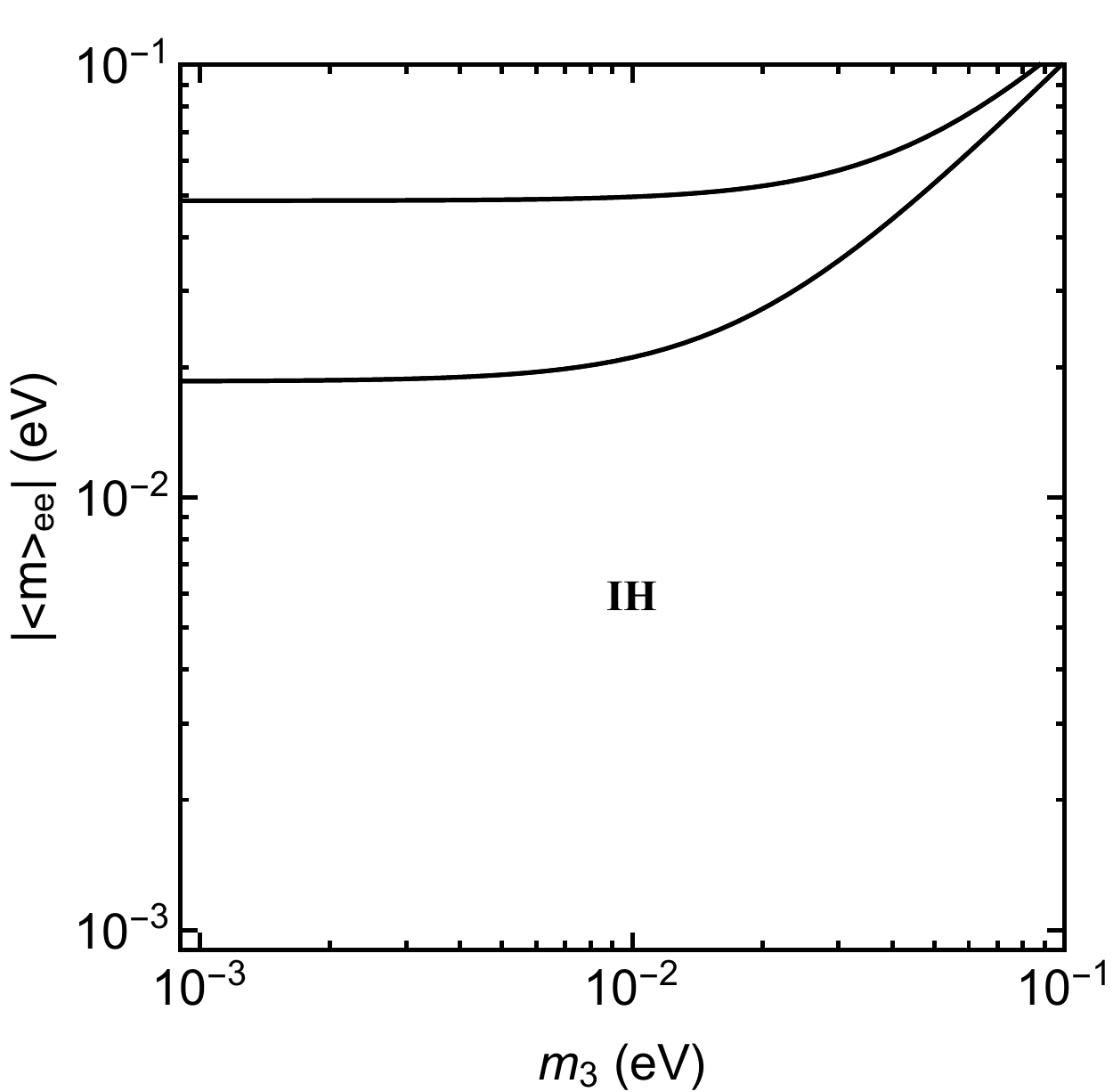}
}
\end{center}
\caption{$\rho$, $\sigma$ and $\left| \langle m \rangle^{}_{ee} \right|$ as functions of the lightest neutrino mass $m^{}_3$ in the case of two sides of equations B, D and E being vanishing.}
\end{figure}

\subsection{A\&B\&C\&F\&G}

In the case of two sides of equations A, B, C, F and G being vanishing, equations AB, AC, AD, AG, BC, BD, CD and EF become ineffective. We are left with no constraint equation. But there are eight new constraint equations which are given by Eqs. (\ref{27}, \ref{28}, \ref{34}) and lead to the following relations for $\bar R^{}_{\alpha \beta} $ and $\bar I^{}_{\alpha \beta}$ (see Eqs. (\ref{30}, \ref{42}, \ref{35}))
\begin{eqnarray}
&& \bar R^{}_{e \mu} = -\bar R^{}_{e \tau} \;,  \hspace{1cm} \bar I^{}_{e \mu} = \bar I^{}_{e \tau} = \bar I^{}_{ee} = \bar I^{}_{\mu \tau} = 0 \;, \nonumber \\
&& \bar I^{}_{\mu \mu} = - \bar I^{}_{\tau \tau} \;, \hspace{1cm} \bar R^{}_{\mu \mu} = \bar R^{}_{\tau \tau} = - \left( \bar R^{}_{ee} + \bar R^{}_{\mu \tau} \right)   \;.
\label{53}
\end{eqnarray}
It is easy to see that these relations can be recombined into
\begin{eqnarray}
&& \bar M^{}_{e\mu} = - \bar M^*_{e\tau} \;, \hspace{1cm} \bar M^{}_{\mu\mu} = \bar M^*_{\tau\tau} \;,
\hspace{1cm} \bar R^{}_{\mu \mu} = \bar R^{}_{\tau \tau} = - \left( \bar R^{}_{ee} + \bar R^{}_{\mu \tau} \right) \;, \nonumber \\
&& \bar M^{}_{ee} \;, \bar M^{}_{e\mu} \;, \bar M^{}_{e\tau} \ {\rm and} \ \bar M^{}_{\mu\tau} \ {\rm being \ real}  \;,
\label{54}
\end{eqnarray}
which suggest that this case can be viewed as a result of $\mu$-$\tau$ reflection symmetry in combination with the condition of $\bar M^{}_{e\mu}$ (and equivalently $\bar M^{}_{e\tau}$) being real if we take $\phi^{}_e = \pi/2$ and $\phi^{}_\mu = - \phi^{}_\tau$.
In total, the number of independent constraint equations  gets increased by three compared to in the general case. So three neutrino mass sum rules will arise.

It turns out that the desired neutrino mass sum rules are given by Eq. (\ref{37}) and $\rho, \sigma = 0$ or $\pi/2$. This can be verified by taking the relations in Eq. (\ref{53}) in the expressions for the neutrino masses in combination with the Majorana CP phases in Eq. (\ref{25}). As discussed in subsection 3.2, these sum rules can only be fulfilled in the IH case. It is found that only for the combination $[\rho, \sigma] = [\pi/2, 0]$ can Eq. (\ref{37}) have a realistic solution $m^{}_3 = 0.021$ eV at which $\left| \langle m \rangle^{}_{ee} \right|$ takes a value of 0.020 eV.

\subsection{A\&B\&D\&E\&G}

In the case of two sides of equations A, B, D, E and G being vanishing, equations AB, AC, AD, AG, BC, BD, CD and EF become ineffective. We are left with no constraint equation. But there are eight new constraint equations given by Eqs. (\ref{27}, \ref{29}, \ref{34}) which lead to the following relations for $\bar R^{}_{\alpha \beta} $ and $\bar I^{}_{\alpha \beta}$ (see Eqs. (\ref{30}, \ref{45}, \ref{35}))
\begin{eqnarray}
&& \bar R^{}_{e \mu} =\bar R^{}_{e \tau} = \bar I^{}_{ee} = 0 \;, \hspace{1cm} \bar I^{}_{e \mu} = - \bar I^{}_{e \tau} \;, \hspace{1cm} \bar R^{}_{ee} = - \bar R^{}_{\mu \tau}  \;, \nonumber \\
&& \bar R^{}_{\mu \mu} = - \bar R^{}_{\tau \tau} \;, \hspace{1cm} \bar I^{}_{\mu \mu} = \bar I^{}_{\tau \tau} = -\bar I^{}_{\mu \tau} \;.
\label{55}
\end{eqnarray}
In total, the number of independent constraint equations  gets increased by three compared to in the general case. So three neutrino mass sum rules will arise.

It turns out that the desired neutrino mass sum rules are given by Eqs. (\ref{32}, \ref{52}). This can be verified by taking the relations in Eq. (\ref{55}) in the expressions for the neutrino masses in combination with the Majorana CP phases in Eq. (\ref{25}). By taking the relations given by Eq. (\ref{52}) in Eq. (\ref{32}), one arrives at a neutrino mass sum rule as
\begin{eqnarray}
\left( m^2_1 - m^2_3 \right) c^4_{12} =  \left( m^2_2 - m^2_3 \right) s^4_{12} \;.
\end{eqnarray}
Unfortunately, this sum rule has no chance to be in agreement with the realistic results.

At this stage, the physical meaning of the various cases we have studied can be clarified in the language of $\mu$-$\tau$ interchange symmetry \cite{MT,review}. This symmetry is defined as follows: In the basis of $M^{}_l$ being diagonal, $M^{}_\nu$ should keep invariant under the $\mu$-$\tau$ interchange operation
and is characterized by
\begin{eqnarray}
M^{}_{e\mu} = M^{}_{e\tau} \;, \hspace{1cm} M^{}_{\mu\mu} = M^{}_{\tau\tau}  \;.
\label{56}
\end{eqnarray}
Such a texture leads to the following predictions for the neutrino mixing parameters
\begin{eqnarray}
\phi^{}_\mu = \phi^{}_\tau + \pi \;, \hspace{1cm} \theta^{}_{23} = \frac{\pi}{4} \;, \hspace{1cm} \theta^{}_{13} = 0  \;,
\label{57}
\end{eqnarray}
implying that
\begin{eqnarray}
\bar M^{}_{e\mu} = - \bar M^{}_{e\tau} \;, \hspace{1cm} \bar M^{}_{\mu\mu} = \bar M^{}_{\tau\tau}  \;.
\label{58}
\end{eqnarray}
Before the measurement of $\theta^{}_{13}$ which was widely believed to be negligibly small at that time, the $\mu$-$\tau$ interchange symmetry was very popular for its predictions. But the observation of a relatively large $\theta^{}_{13}$ \cite{dayabay} forces us to consider breaking of this symmetry. Nevertheless, we might have a partial $\mu$-$\tau$ interchange symmetry where part of the following four relations given by Eq. (\ref{58}) still hold
\begin{eqnarray}
\bar R^{}_{e\mu} = - \bar R^{}_{e\tau} \;, \hspace{1cm} \bar I^{}_{e\mu} = - \bar I^{}_{e\tau} \;, \hspace{1cm} \bar R^{}_{\mu\mu} = \bar R^{}_{\tau\tau}  \;, \hspace{1cm} \bar I^{}_{\mu\mu} = \bar I^{}_{\tau\tau}  \;.
\label{59}
\end{eqnarray}
From our results, it is found that: (1) In the case of two sides of equations A and G being vanishing, the relation $\bar R^{}_{e\mu} = - \bar R^{}_{e\tau}$ still holds. (2) In the case of two sides of equation B being vanishing, the relation $\bar I^{}_{e\mu} = - \bar I^{}_{e\tau}$ still holds. (3) In the case of two sides of equations C and F being vanishing, the relation $\bar R^{}_{\mu\mu} = \bar R^{}_{\tau\tau}$ still holds. (4) In the case of two sides of equations D and E being vanishing, the relation $\bar I^{}_{\mu\mu} = \bar I^{}_{\tau\tau}$ still holds. (5) In the case of two sides of equations A, B and G being vanishing, the relations $\bar R^{}_{e\mu} = - \bar R^{}_{e\tau}$ and $\bar I^{}_{e\mu} = - \bar I^{}_{e\tau}$ which can be recombined into $\bar M^{}_{e\mu} = - \bar M^{}_{e\tau}$ still hold. (6) In the case of two sides of equations A, C, F and G being vanishing, the relations $\bar R^{}_{e\mu} = - \bar R^{}_{e\tau}$ and $\bar R^{}_{\mu\mu} = \bar R^{}_{\tau\tau}$ still hold. In this case, one can say that the real part of $\bar M^{}_\nu$ still respects the $\mu$-$\tau$ interchange symmetry. (7) In the case of two sides of equations A, D, E and G being vanishing, the relation $\bar I^{}_{\mu\mu} = \bar I^{}_{\tau\tau}$ still holds, while $\bar R^{}_{e\mu}$ and $\bar R^{}_{e\tau}$ vanish. (8) In the case of two sides of equations B, C and F being vanishing, the relation $\bar R^{}_{\mu\mu} = \bar R^{}_{\tau\tau}$ still holds, while $\bar I^{}_{e\mu}$ and $\bar I^{}_{e\tau}$ vanish. (9) In the case of two sides of equations B, D and E being vanishing, the relations $\bar I^{}_{e\mu} = - \bar I^{}_{e\tau}$ and $\bar I^{}_{\mu\mu} = \bar I^{}_{\tau\tau}$ still hold. In this case, one can say that the imaginary part of $\bar M^{}_\nu$ still respects the $\mu$-$\tau$ interchange symmetry. (10) In the case of two sides of equations A, B, C, F and G being vanishing, the relations $\bar R^{}_{e\mu} = - \bar R^{}_{e\tau}$ and $\bar R^{}_{\mu\mu} = \bar R^{}_{\tau\tau}$ still hold, while $\bar I^{}_{e\mu}$ and $\bar I^{}_{e\tau}$ vanish. In this case, one can say that the real part of $\bar M^{}_\nu$ still respects the $\mu$-$\tau$ interchange symmetry. (11) In the case of two sides of equations A, B, D, E and G being vanishing, the relations $\bar I^{}_{e\mu} = - \bar I^{}_{e\tau}$ and $\bar I^{}_{\mu\mu} = \bar I^{}_{\tau\tau}$ still hold, while $\bar R^{}_{e\mu}$ and $\bar R^{}_{e\tau}$ vanish. In this case, one can say that the imaginary part of $\bar M^{}_\nu$ still respects the $\mu$-$\tau$ interchange symmetry.

\section{Summary}

Motivated by the fact that the current neutrino oscillation data is consistent with maximal atmospheric mixing angle and Dirac CP phase, we derive in a novel approach the possible textures of neutrino mass matrix that can lead us to $\theta^{}_{23} = \pi/4$ and $\delta = - \pi/2$. In order to evade the uncertainties created by the unphysical phases, we work on the effective neutrino mass matrix $\bar M^{}_\nu$ instead of $M^{}_\nu$ itself. Since the unphysical phases have cancelled out in $\bar M^{}_\nu$, its twelve components $\bar R^{}_{\alpha \beta}$ and
$\bar I^{}_{\alpha \beta}$ are not all independent but subject to three constraint equations. After imposing the conditions $\theta^{}_{23} = \pi/4$ and $\delta = - \pi/2$, there are five independent constraint equations for $\bar R^{}_{\alpha \beta}$ and $\bar I^{}_{\alpha \beta}$. We derive these constraint equations (i.e., Eqs. (\ref{20}, \ref{24}) and three independent ones from Eq. (\ref{22})) by eliminating $\theta^{}_{12}$ and $\theta^{}_{13}$ in Eq. (\ref{16}) in the general case where none of equations A-G has its two sides vanish. On the basis of this, we further study the possible textures of $\bar M^{}_\nu$ by considering that some of equations A-G may have their two sides vanish.
When an equation has its two sides vanish, the constraint equation(s) resulting from it will become ineffective. But the fact that its two sides vanish itself brings about two new constraint equations. So the number of independent constraint equations gets increased compared to in the general case. When this number gets increased by one (and so on), there will correspondingly be one (and so on) neutrino mass sum rules relating the neutrino masses and Majorana CP phases.

Thanks to the observations that equations A and G (or C and F or D and E) always have their two sides vanish simultaneously and equations E and F (or A, B, C, D and G) are not allowed to have their two sides vanish simultaneously, one just needs to consider the cases where
equations A\&G, B, C\&F, D\&E, A\&B\&G, A\&C\&F\&G, A\&D\&E\&G, B\&C\&F, B\&D\&E, A\&B\&C\&F\&G or A\&B\&D\&E\&G have their two sides vanish. In the case of two sides of equations A\&G, B, C\&F or D\&E being vanishing, there is one neutrino mass sum rule. In the case of two sides of equations A\&B\&G, A\&C\&F\&G, A\&D\&E\&G, B\&C\&F or B\&D\&E being vanishing, there are two neutrino mass sum rules. In the case of two sides of A\&B\&C\&F\&G or A\&B\&D\&E\&G being vanishing, there are three neutrino mass sum rules. The neutrino mass sum rule Eq. (\ref{37}) arising from the vanishing of two sides of equation B can only be fulfilled in the IH case. By taking $\phi^{}_e =\pi/2$ and $\phi^{}_\mu = - \phi^{}_\tau$, the texture of $\bar M^{}_\nu$ obtained in the case of two sides of equations A\&C\&F\&G being vanishing can reproduce the specific texture given by the neutrino $\mu$-$\tau$ reflection symmetry. In the case of two sides of equations A\&B\&C\&F\&G being vanishing, the unknown neutrino parameters can be completely determined: the neutrino masses are of the inverted hierarchy with $m^{}_3 = 0.021$ eV while the Majorana CP phases are $[\rho, \sigma] = [\pi/2, 0]$. But in the case of two sides of equations A\&B\&D\&E\&G being vanishing, the resulting neutrino mass sum rules have no chance to be in agreement with the experimental results. As discussed at the end of section 3, the various cases we have studied can find a motivation from the partial $\mu$-$\tau$ interchange symmetry.

Finally, we point out that the results obtained in this work can be further studied from two aspects: On the one hand, one can study the origins of these special textures from some underlying flavor symmetries in the lepton sector \cite{review2}. On the other hand, one can study the breaking effects of these special textures so as to accommodate the deviations of $\theta^{}_{23}$ and $\delta$ from $\pi/4$ and $-\pi/2$.

\vspace{0.5cm}

\underline{Acknowledgments} \hspace{0.2cm} This work is supported
in part by the National Natural Science Foundation of China under grant Nos. 11747318, 11875157 (Z.C.L. and C.X.Y.) and 11605081 (Z.H.Z.).

\end{document}